\documentclass[11pt]{article}
\usepackage[utf8]{inputenc}
\usepackage{setspace,geometry}
\usepackage{graphicx}
\usepackage[dvipsnames]{xcolor}
\usepackage{colortbl}
\usepackage{subfig}
\usepackage{mwe}
\usepackage{soul}
\usepackage{amsmath,amssymb}
\usepackage{bbm,bm}
\usepackage[hidelinks]{hyperref}
\usepackage{natbib,authblk}
\setcitestyle{authoryear}
\setcitestyle{authoryear,open={(},close={)}}
\usepackage{multirow}
\usepackage{booktabs}
\usepackage{dcolumn}
\usepackage{enumitem}
\usepackage{tikz}
\usetikzlibrary{"automata","positioning"}
\usetikzlibrary{fit}
\usetikzlibrary{calc}
\usetikzlibrary{decorations.pathreplacing,tikzmark}
\usetikzlibrary{arrows.meta,angles,quotes}
\tikzset{>={Latex[width=1.5mm,length=1.5mm]}}

\newcommand{\xoverbrace}[2][\vphantom{\mathbf{Z(\mathbf{x})}^\top}]{\overbrace{#1#2}}

\newcommand{\myfrac}[3][0pt]{\genfrac{}{}{}{}{\raisebox{#1}{$#2$}}{\raisebox{-#1}{$#3$}}}

 \definecolor{shadecolor}{rgb}{0.75,0.75,0.75}

 \newcolumntype{d}[1]{D{.}{.}{#1}}

\title{How to account for behavioural states in step-selection analysis: a model comparison}

\author{
J.\ Pohle$^{1}$\footnote{
Corresponding author; email: \texttt{jennifer.pohle@uni-potsdam.de}; postal address: Am Mühlenberg 3, building 60, 14476 Potsdam, Germany.
}, 
J.\ Signer$^{2}$,
J.\ A.\ Eccard$^{3}$,
M.\ Dammhahn$^{4}$ and
U.\ E.\ Schlägel$^{1}$\\
$^1$Institute of Biochemistry and Biology, University of Potsdam, Potsdam, Germany \\ $^2$Wildlife Sciences; Faculty of Forest Sciences and Forest Ecology; University of Goettingen, G\"ottingen, Germany \\
$^3$Animal Ecology, University of Potsdam, Potsdam, Germany \\
$^4$ Behavioural Biology, University of M\"unster, M\"unster, Germany
}
\date{}

\begin{document}

\begin{spacing}{1.2}
\maketitle
\vspace{-2em}

\begin{abstract}
\noindent
\begin{enumerate}
    \item Step-selection models are widely used to study animals' fine-scale habitat selection based on movement data. Resource preferences and movement patterns, however, can depend on the animal's unobserved behavioural states, such as resting or foraging. This is ignored in standard (integrated) step-selection analyses (SSA, iSSA). While different approaches have emerged to account for such states in the analysis, the performance of such approaches and the consequences of ignoring the states in the analysis have rarely been quantified.
    \item We evaluated the recent idea of combining hidden Markov chains and iSSA in a single modelling framework. The resulting Markov-switching integrated step-selection analysis (MS-iSSA) allows for a joint estimation of both the underlying behavioural states and the associated state-dependent habitat selection. In an extensive simulation study, we compared the MS-iSSA to both the standard iSSA and a classification-based iSSA (i.e., a two-step approach based on a separate prior state classification). We further compared the three approaches in a case study on fine-scale interactions of simultaneously tracked bank voles (\textit{Myodes glareolus}).
    \item The simulation study illustrates that standard iSSAs lead to erroneous conclusions due to both biased estimates and unreliable p-values when ignoring underlying behavioural states. We found the same for iSSAs based on prior state-classifications, as they ignore misclassifications and classification uncertainties. The MS-iSSA, on the other hand, performed well in parameter estimation and decoding of behavioural states. In the bank-vole case study, the MS-iSSA was able to distinguish between an inactive and active state, but results highly varied between individuals.
    \item MS-iSSA provides a flexible framework to study state-dependent habitat selection. It defines states on both selection and movement patterns and accounts for uncertainties in the corresponding state process. To facilitate its use, we implemented the MS-iSSA approach in the R package \textit{msissa}.
\end{enumerate}
\end{abstract}
\vspace{0.5em}


\section{Introduction}\label{Sec_Intro}

Combining animal movement and environmental data, step-selection analysis (SSA) and its extension, the integrated step-selection analysis (iSSA), build a popular framework for studying animals' fine-scale habitat selection, while also taking the movement capacity of the animal into account \citep{for05,for09,avg16,nor21}. Essentially, SSA and iSSA are used to explain the animals' space use based on possible preferences for or avoidance of environmental features, accounting for spatial limitations that the animals' movement process imposes on availability. ISSA has successfully been applied, for example, to analyse elk response to roads \citep{pro17}, to study the effects of artificial nightlight on predator–prey dynamics of cougars and deer \citep{dit21}, and to model space use of Cape vultures in the context of wind energy development \citep{cer23}. Besides conventional habitat use, SSA has also proven suitable for detecting interactions such as avoidance or attraction between simultaneously tracked individuals \citep{schl19}. 

For parameter estimation, SSA and iSSA use a conditional logistic regression for case-control designs to compare the characteristics of observed, i.e.\ \textit{used} steps against the covariates of alternative steps \textit{available} at a given time point. In this context, a step is the straight-line segment connecting two consecutive locations sampled at regular time intervals and is usually described by the step length and turning angle, i.e.\ the directional change \citep{for05}. The covariates usually correspond to features of the steps' end point, e.g.\ vegetation or snow cover \citep{str21}, but can also refer to characteristics along the step, e.g.\ the presence of roads on the path \citep{pro17}. What is considered to be available at a given time point depends on the assumptions made about the animals' movement capacities and/or typical, i.e.\ habitat-selection-free movement patterns. This usually translates to assumptions about the animals' step length and turning angle distribution (e.g.\ gamma and von Mises distributions). Tentative parameter estimates for these distributions can be estimated from observed steps. These estimates are biased because movement is restricted by habitat selection. A correction for the tentative parameters can be estimated using iSSA \citep{avg16,fie21}.

\begin{figure}[!t]
	\centering
    	\includegraphics[width=0.95\textwidth]{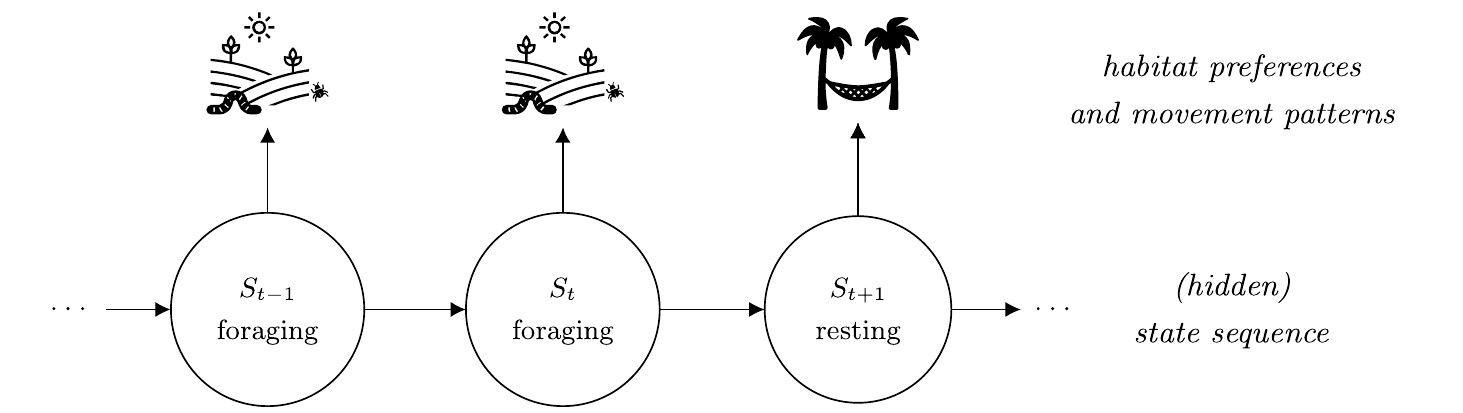}
	\caption{Illustration of how behavioural states can affect animals habitat selection and movement patterns. The state ``foraging'' is related to search for food such as small insects in an open landscape, while the state ``resting'' is associated to a retreat in its shelter. Usually, the behavioural states are unobserved, thus hidden, and serially correlated. This structure corresponds to the basic dependence structure of a Markov-switching step-selection model.}
	\label{fig:HMM}
\end{figure}

While (i)SSAs seem suitable in numerous instances, it has recently been argued that fine-scale habitat selection, resource requirements, and selection-free movement patterns might depend on the animal's behavioural modes such as resting or foraging (illustrated in Figure \ref{fig:HMM}). Ignoring such states in the analysis might thus lead to biased results and misleading conclusions \citep{roe14,sur19}. With telemetry-based location data, however, the underlying behavioural states are usually unobserved. Therefore, it has been suggested to first classify the movement data into different states, e.g.\ based on hidden Markov models (HMMs, \citealp{zuc16}), and to split the step observations accordingly into state-specific data sets, which can then be used to fit state-specific (i)SSAs in a second step \citep{roe14,kar19,pic22}. This two-step approach, hereafter named TS-(i)SSA, accounts for the unobserved state structure and is convenient as it can be based on existing software implementations. It has, however, two major drawbacks. First, the state classification is purely based on movement patterns without considering habitat selection. Thus, habitat selection and selection-independent movement processes can be confounded when defining the states. This can affect the validity of the state classification and can lead to a bias in the estimated movement and selection parameters \citep{pri22}. Second, the uncertainty in the HMM state classification is completely ignored in the follow-up (i)SSA. This can again lead to biased movement and (habitat) selection coefficients, as misclassification can occur. Furthermore, confidence intervals and standard p-values are no longer reliable as the uncertainty of both the HMM parameter estimation and the state classification are not taken into account. Consequently, also the TS-iSSA might lead to biased results and misleading conclusions. How serious this is in practice, however, has rarely been quantified. \citet{pri22} evaluated a population-level version of the TS-iSSA in a simulation study and found good classification and prediction performances in the scenarios considered, but also biased parameter estimates. However, as they focused on the population level, they did not provide results on the variation, uncertainty quantification and estimation accuracy of the individually fitted TS-iSSA models.

The above mentioned problems could be avoided by combining step selection models and hidden Markov chains (as used in HMMs) in a single model to allow for a joint estimation of the underlying state, habitat selection and selection-free movement processes. Similar as \citet{nic17} and \citet{pri22}, we therefore consider a Markov-switching integrated step-selection analysis (MS-iSSA, also called HMM-SSA) which renders a prior state classification unnecessary. All model parameters are jointly estimated using a case-control Markov-switching conditional logistic regression framework. In our implementation, we use a numerical maximum likelihood estimation and constrain all parameters to their natural parameter space to avoid problems in model interpretation (e.g.\ a negative shape parameter for an assumed gamma distribution for step length). For state decoding, we consider the well-known Viterbi algorithm, which computes the most likely state sequence underlying the data given the fitted model.

The aim of this paper is two-fold. First, we provide a broad overview of the MS-iSSA framework by discussing the underlying movement model, its relation to alternative approaches (iSSA, TS-iSSA, HMMs) and, most importantly, its practical implementation, which we further facilitate through release as an R package. Second, we investigate whether and to what extent either the complete neglect of underlying states in the analysis (iSSA) or their incorporation by a prior HMM-based state classification (TS-iSSA) affects the estimation results compared to the MS-iSSA approach. For this, we use an extensive simulation study to compare the estimation and, if applicable, classification performance of iSSAs, TS-iSSAs and MS-iSSAs in three state-switching scenarios. Thereby, we showcase different ways in which behavioural states could influence the animals' movement decisions. A supplementary simulation covers a scenario without underlying state-switching. We further compare the three approaches in a case study on fine-scale interactions of bank voles (\textit{Myodes glareolus}), which are small ground-dwelling rodents. Using a movement data set of synchronously tracked individuals, as analysed in \citet{schl19}, we test whether MS-iSSAs can detect meaningful biological states and whether they provide new insights into interactions, such as attraction, avoidance, or neutrality towards other conspecifics compared to iSSAs.


\section{Methods}\label{Sec2_Methods}


\subsection{Markov-switching step-selection model}\label{Sec2_basics}

We use $\{\mathbf{x}_{0,1},\mathbf{x}_{0,2},\ldots,\mathbf{x}_{0,T}\}$ to denote the sequence of two-dimensional animal locations observed at regular time intervals, which forms the observed movement track. Conditional on the previous location $\mathbf{x}_{0,t-1}$, a step from the current location $\mathbf{x}_{0,t}$ to the next location $\mathbf{x}_{0,t+1}$, is characterised by its step length $l_{0,t+1}$, i.e.\ the straight-line distance between the two consecutive locations, and its turning angle $\alpha_{0,t+1}$, i.e.\ the directional change. The corresponding covariate vector $\mathbf{Z}_{0,t+1}$ stores the feature values of the step, and we use $\mathbf{Z}$ to denote the collection of covariate values for all possible location in the given area.

In the Markov-switching step-selection model, we assume the observed steps to be driven by an underlying hidden state sequence $\{S_{1},S_{2},\ldots,S_{T}\}$ with $N$ discrete states. Thus, each state variable $S_t$ at time $t$ can take one of $N$ state values ($S_t \in \{1,\ldots,N\}$). These states serve as proxies for the unobserved behavioural modes of the animal that influence its movement and habitat selection (illustrated in Figure \ref{fig:HMM}). We assume the state sequence to be a homogeneous $N$-state Markov chain, characterised by its transition probabilities $\gamma_{ij}=\Pr(S_t=j\mid S_{t-1}=i)$ to switch from state $i$ to state $j$, summarised in the $N \times N$ transition probability matrix $\bm{\Gamma}$, and the initial state distribution $\bm{\delta}$ which contains the probabilities to start in a certain state.

Each state $i$ ($i=1,\ldots,N$) is associated to a state-dependent density $f_i$ generating the next location. Its functional form is similar to the basic step-selection model \citep{for09}, but with movement and habitat selection parameters being state-dependent. Thus, conditional on locations $\mathbf{x}_{0,t-1}$ and $\mathbf{x}_{0,t}$, and covariates $\mathbf{Z}$, the current state $S_t=i$ determines the following distribution for a step to location $\mathbf{x}_{0,t+1}$:

\begin{align*}
    f_i(\mathbf{x}_{0,t+1} \mid \mathbf{x}_{0,t}, \mathbf{x}_{0,t-1}, \mathbf{Z}; \boldsymbol{\theta}_i, \boldsymbol{\beta}_i)&=
        \cfrac{\xoverbrace{\phi(\mathbf{x}_{0,t+1} \mid \mathbf{x}_{0,t},\mathbf{x}_{0,t-1};\boldsymbol{\theta}_i)}
    ^{\substack{\text{selection-free}\\\text{movement kernel}}}
    \cdot
    \overbrace{\omega(\mathbf{Z}_{0,t+1};\boldsymbol{\beta}_i)}
    ^{\substack{\text{movement-free}\\\text{selection function}}}}
    {\underbrace{\int_{\mathbf{\tilde{x}} \in \mathcal{D}}
    \phi(\mathbf{\tilde{x}} \mid \mathbf{x}_{0,t},\mathbf{x}_{0,t-1};\boldsymbol{\theta}_i)
    \cdot
    \omega(\mathbf{\tilde{Z}};\boldsymbol{\beta}_i) 
    d \mathbf{\tilde{x}}}
    _{\text{normalising constant}}}
\end{align*}

The density consists of three components: i) The movement kernel $\Phi(\cdot)$ describes the space use in a homogeneous landscape and is usually defined in terms of step length  $l_{0,t+1}$ (e.g.\ gamma distribution) and turning angle $\alpha_{0,t+1}$ (e.g.\ von Mises distribution). The corresponding state-dependent parameters for state $i$ are summarised in the movement parameter vector $\boldsymbol{\theta}_i$; ii) The movement kernel is weighted by the movement-free selection function $\omega(\cdot)$ which indicates a possible selection for or against the covariates in $\mathbf{Z}_{0,t+1}$. It is usually assumed to be a log-linear function of the (state-dependent) selection coefficient vector $\boldsymbol{\beta}_i$, 

$$\omega(\mathbf{Z}_{0,t+1};\boldsymbol{\beta}_i)=\exp\left( \mathbf{Z}_{0,t+1}^{\top}\boldsymbol{\beta}_i \right),$$

where a positive selection coefficient value indicates preference for, and a negative value avoidance of a corresponding covariate; iii) The integral in the denominator ensures that $f_i$ integrates to one. Usually, it is analytically intractable and must therefore be approximated, for example, using numerical integration methods. We provide an example of state-dependent step-selection densities in a $2$-state scenario in Figure S1 (Supplementary Material).

There are important relations between the Markov-switching step-selection model and two alternative movement models: i) If all states share the same parameters, i.e.\ $\boldsymbol{\theta}_1=\ldots=\boldsymbol{\theta}_N$ and $\boldsymbol{\beta}_1=\ldots=\boldsymbol{\beta}_N$, or if the number of states is set to one, i.e.\ $N=1$, the model reduces to the basic step-selection model without state-switching \citep{for09}; ii) If all selection coefficients are equal to zero, i.e.\ $\boldsymbol{\beta}_1=\ldots=\boldsymbol{\beta}_N=\mathbf{0}$, the model reduces to a basic movement HMM (\citealp{lan12}, \citealp{pat17}) with state-dependent step length and turning angle distributions as implied by the movement kernel $\phi(\cdot)$ but without habitat selection. These relations are very convenient, for example in the context of model comparison and model selection, as it allows the use of standard tests or information criteria to select between these three candidate models.

We can simplify the step-selection density $f_i$ by assuming step length to follow a distribution from the exponential family (with support on non-negative real numbers, e.g.\ a gamma distribution) and turning angle to follow either a uniform or von Mises distribution with fixed mean \citep{avg16,nic17}. In this case, the product of the movement kernel $\phi(\cdot)$ and the exponential selection function $\omega(\cdot)$ is proportional to a single log-linear function of the corresponding model parameters and $f_i$ reduces to: 

\begin{align*}
    f_i(\mathbf{x}_{0,t+1} \mid \mathbf{x}_{0,t}, \mathbf{x}_{0,t-1}, \mathbf{Z}; \boldsymbol{\theta}_i, \boldsymbol{\beta}_i) &=\myfrac[3pt]{\exp\left(\mathbf{C}_{0,t+1}^{\top} \boldsymbol{\theta}_i + \mathbf{Z}_{0,t+1}^{\top}\boldsymbol{\beta}_i-\log(l_{0,t+1})\right)}
    {\displaystyle \int_{\mathbf{\tilde{x}} \in \mathcal{D}}
    \exp\left(\mathbf{\tilde{C}}^{\top} \boldsymbol{\theta}_i + \mathbf{\tilde{Z}}^{\top}\boldsymbol{\beta}_i-\log(\tilde{l})\right)
    d \mathbf{\tilde{x}}}.
\end{align*}

The vector $\mathbf{C}_{0,t+1}$ can be interpreted as a movement covariate vector that contains different step length and turning angle terms. Its exact form depends on the chosen step length and turning angle distributions (Table S1, see also \citealp{nic17}). For example, for gamma-distributed step length and von-Mises-distributed turning angles with mean zero, we have $\mathbf{C}_{0,t+1} = ( \log(l_{0,t+1}), -l_{0,t+1},\allowbreak \cos(\alpha_{0,t+1}))^{\top}$. The corresponding state-dependent movement coefficient vector is $\boldsymbol{\theta}_i=(k_i-1,r_i,\kappa_i)^{\top}$ with $k_i$ and $r_i$ being the shape and rate parameter of the gamma-distribution belonging to state $i$, respectively, and $\kappa_i$ being the state-dependent concentration parameter of the von-Mises distribution. Thus, in this reduced representation of the step-selection density $f_i$, the parameterisation of the movement kernel might differ from the commonly used parameterisation of the corresponding step and angle distributions (e.g.\ $k_i-1$ instead of $k_i$), but there is a direct relationship between the two (Table S1 and S2). The negative log step length included in the exponential function is necessary to correctly represent the movement kernel in a Cartesian coordinate system. 

The reduced form of $f_i$ is very convenient. Justified by the law of large numbers, it allows for a joint parameter estimation of the state, movement and selection parameters based on a Markov-switching conditional logistic regression for case-control designs with $M$ control, i.e.\ available, locations per observed, i.e.\ used, location (\citealp{nic17}, see also \citealt{avg16} for step-selection models without state-switching). This forms the basis for the MS-iSSA.


\subsection{Markov-switching integrated step-selection analysis}\label{Sec2_est}

 The MS-iSSA workflow is similar to the one of the iSSA. For each observed step, we choose $M$ control steps, e.g.\ using a suitable proposal distribution for step length and turning angle, respectively, and extract the corresponding habitat and movement covariate values. This builds the case-control data set. The model parameters are then estimated using a Markov-switching conditional logistic regression, i.e.\ a conditional logistic regression in which the regression coefficients depend on an underlying latent Markov chain. In our implementation, we use the forward algorithm, which is well-known especially in the context of HMMs \citep{zuc16} to efficiently evaluate the corresponding likelihood. This allows for a numerical maximum likelihood estimation based on standard optimisation procedures such as \textit{nlm} in R \citep{RCore22}. Afterwards, it is possible to decode the states, for example, using the Viterbi algorithm \citep{vit67}, which calculates the most likely sequence of states given the fitted model and the case-control data.

 More precisely, for each step from location $\mathbf{x}_{0,t}$ to $\mathbf{x}_{0,t+1}$ ($t=2,\ldots,T-1$), we have a choice set $\tilde{\mathbf{x}}_{t+1}=\{\mathbf{x}_{0,t+1},\mathbf{x}_{1,t+1},\ldots,\mathbf{x}_{M,t+1}\}$ that includes the observed and the $M$ control locations for the end point of the step. Usually, the control locations are randomly drawn from a suitable proposal distribution for step length and turning angle \citep{for09}. However, it is also possible to use a grid or a mesh \citep{arc23}. Here the devil is in the detail, as depending on the sampling procedure, the interpretation of the models' movement coefficients might differ (see Section S2 in the Supplementary Material). The interpretation of the selection coefficients, however, remain unaffected.

In the Markov-switching conditional logistic regression, we model the state-dependent choice probability $p_{0ti}$ of choosing the observed location $\mathbf{x}_{0,t+1}$ from the choice set $\tilde{\mathbf{x}}_{t+1}$ given the current state $S_t=i$, as:

\begin{align*}
    p_{0ti}(\mathbf{x}_{0,t+1}|\tilde{\mathbf{x}}_{t+1},\mathbf{C},\mathbf{Z};\boldsymbol{\theta}_i,\boldsymbol{\beta}_i)=
    \myfrac[2pt]{\exp\left(\mathbf{C}_{0,t+1}^{\top}\boldsymbol{\theta}_i +  \mathbf{Z}_{0,t+1}^\top\boldsymbol{\beta}_i\right)}
    {\sum_{m=0}^M 
    \exp\left(\mathbf{C}_{m,t+1}^{\top}\boldsymbol{\theta}_i +  \mathbf{Z}_{m,t+1}^\top\boldsymbol{\beta}_i\right)},
\end{align*}

with $\mathbf{C}_{m,t+1}$ and $\mathbf{Z}_{m,t+1}$ being the movement and habitat covariate vectors belonging to location $\mathbf{x}_{m,t+1}$ for $m=0,\ldots,M$. This case-control step-selection probability is closely related to direct numerical integration, which offers an alternative way to approximate the step-selection density $f_i$. We derive the likelihood of the Markov-switching conditional logistic regression by plugging $p_{0ti}$ ($i=1,\ldots,N$) into the HMM likelihood \citep{zuc16},

\begin{align*}
	\mathcal{L}(\boldsymbol{\theta},\boldsymbol{\beta};\tilde{\mathbf{x}}_3,\tilde{\mathbf{x}}_4,\ldots,\tilde{\mathbf{x}}_{T},\mathbf{C},\mathbf{Z})
	&= \boldsymbol{\delta}^\top \mathbf{P}(\tilde{\mathbf{x}}_{3}) \bm{\Gamma} \mathbf{P}(\tilde{\mathbf{x}}_{4}) \bm{\Gamma} \cdots \boldsymbol{\Gamma} \mathbf{P}(\tilde{\mathbf{x}}_{T}) \bm{1},
\end{align*}

where $\mathbf{P}(\tilde{\mathbf{x}}_{t})=\text{diag}(p_{0t1},\ldots,p_{0tN})$ is a diagonal matrix including the state-dependent step-selection probabilities, $\bm{\Gamma}$ and $\boldsymbol{\delta}$ are the transition probability matrix and the initial distribution of the underlying Markov chain, respectively, and $\mathbf{1}$ is an $N$-dimensional vector of ones. We can then estimate the model parameters using a numerical maximisation of the log-likelihood (for details, see \citealp{zuc16}). In our implementation, we restrict the movement parameters to always remain in their natural parameter space, e.g.\ the shape and rate parameters of the gamma distribution are always greater than zero.

For initialisation, the numerical maximisation requires a set of starting values for the model parameters. To avoid ending up in a local maximum of the log-likelihood, it is necessary to test several sets of starting values, for example by randomly drawing values for each model parameter. We discuss this in more detail in Section S3 in the Supplementary Material.


\subsection{Two-step approach}\label{Sec2_TSiSSA}

The TS-iSSA is based on the same idea as the MS-iSSA. However, the TS-iSSA relies on a \textit{prior} classification of the movement data into different movement states. Thus, in a first step, an $N$-state HMM with state-dependent step length and turning angle distributions as defined for the movement kernel is fitted to the data, e.g.\ using a gamma distribution for step length and a von Mises distribution for turning angles. Then, the Viterbi algorithm is used to assign each observed step to one of the $N$ HMM movement states. Alternatively, local state decoding can be used. In the second step, state-specific (i)SSAs are fitted to the state-specific data using a case-control design and conditional logistic regression (e.g.\ \citealp{roe14,kar19}). The control steps for the state-specific case-control data sets are thereby sampled based on the respective state-dependent HMM step length and turning angle distributions.


\subsection{Simulation Study}\label{Sec_Sim}

\begin{figure}
    \centering
    \includegraphics[width=0.35\textwidth]{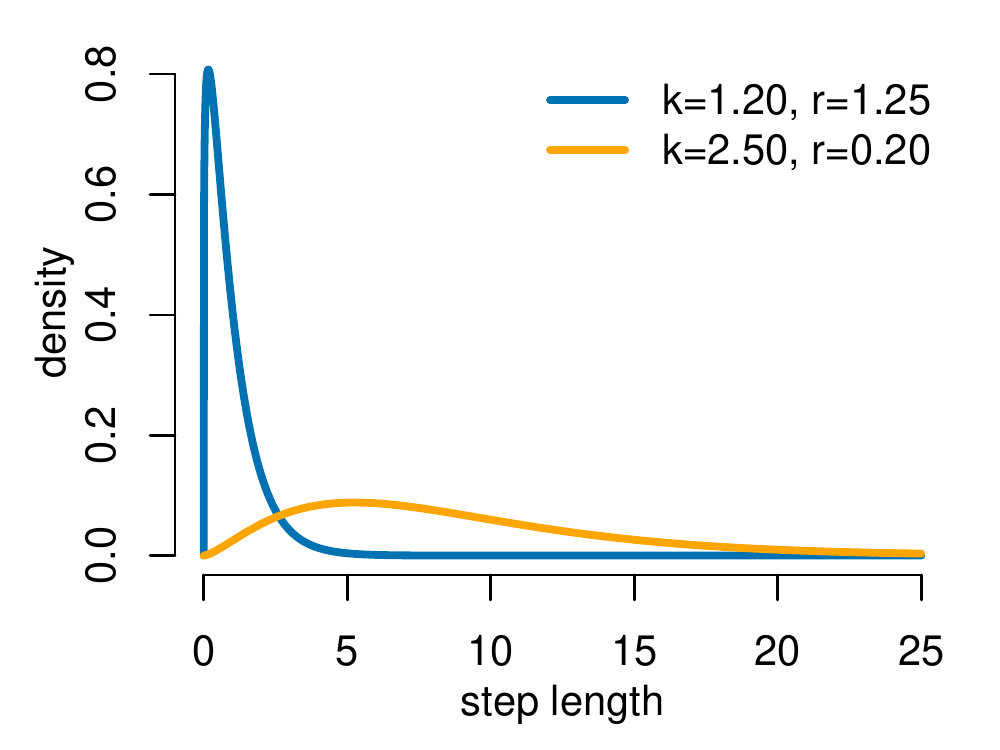}
    \includegraphics[width=0.35\textwidth]{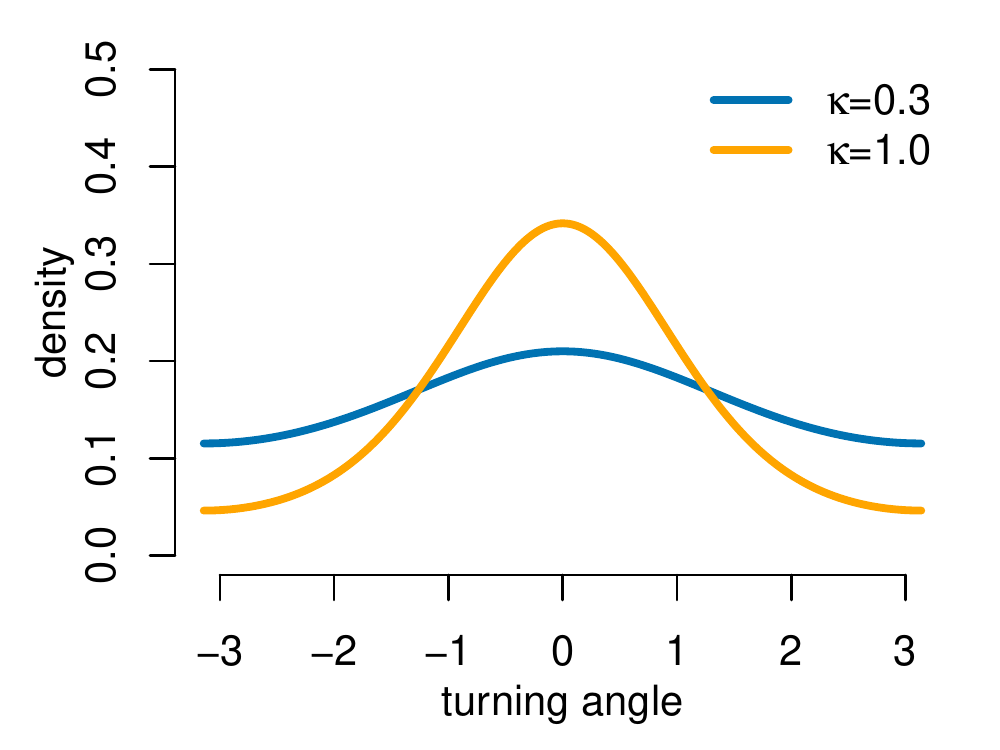}
    \caption{Gamma and von Mises distributions for step length and turning angle, respectively, which are used in the simulation study to form the state-dependent movement kernels of the two states. The corresponding parameters are denoted by $k$ (shape), $r$ (rate) and $\kappa$ (concentration). The distributions of state $2$ are shown in orange. The distributions for state $1$ depend on the simulation scenario: In Scenario 2, both states share the same movement kernel and thus, the distributions in orange are also used for state $1$; in the other two Scenarios 1 and 3, the distributions in blue are used for state $1$.}
    \label{fig:sim_sdds}
\end{figure}

We used a simulation study with three state-switching scenarios to evaluate the performance of our MS-iSSA approach and to demonstrate possible consequences of either ignoring the underlying latent states in the traditional iSSA or ignoring the uncertainty of prior state-decoding in the TS-iSSA. For each scenario, we generated movement data from a Markov-switching step-selection model with $2$ states and state transition probabilities $\gamma_{11}=\gamma_{22}=0.9$. A realisation of a Gaussian random field with covariance $\sigma^2=1$ and range $\phi=10$, computed using the function \textit{grf} from the R-package \textit{geoR} \citep{rib22}, served as the habitat covariate $\mathbf{Z}$ (Figure S2). For the movement kernel, we used state-dependent gamma and zero-mean von Mises distributions to model step length and turning angle, respectively (Figure \ref{fig:sim_sdds}). 

Table \ref{tab:sim_sen} summarises the movement and selection parameters for each of the three simulation scenarios. Scenario 1 is chosen to represent a typical inactive-active scenario in which the first state (``inactive'' state) is associated to small step length, less directive movement and no selection, while the second state (``active'' state) corresponds to larger step length, more directed movement and attraction to the landscape feature $\mathbf{Z}$. The second and the third scenarios cover the rather extreme cases in which either the selection or the movement parameters are shared across states: In Scenario 2 (``switching preferences''), the two states only differ in their selection patterns with avoidance of the feature in state 1, and attraction to the feature in state 2. In Scenario 3 (``HMM''), only the movement patterns differ across states while there is no selection for or against the landscape feature in either state. This corresponds to a basic movement HMM (see Section \ref{Sec2_basics}). To check the robustness of the MS-iSSA, in Section S5.2 of the Supplementary Material we additionally cover a fourth scenario without state-switching in which the data are generated based on a standard step-selection model. Furthermore, to check the influence of the spatial variation in the habitat feature on the estimation results, we also considered a second landscape feature map which was a realisation of a Gaussian random field with covariance $\sigma^2=1$ and range $\phi=50$ (Section S5.3 in the Supplementary Material).

\begin{table}[t]
    \centering
    \footnotesize
    \begin{tabular}{ld{2.2}rrrd{2.2}rrr}
        \toprule
         & \multicolumn{4}{c}{state 1} & \multicolumn{4}{c}{state 2} \\\cmidrule(lr{.75em}){2-5}\cmidrule(lr{.75em}){6-9}
         & \multicolumn{1}{c}{select.\ fun.\ } &  \multicolumn{3}{c}{movement kernel} & \multicolumn{1}{c}{select.\ fun.\ } &  \multicolumn{3}{c}{movement kernel} \\\cmidrule(lr{.75em}){2-2}\cmidrule(lr{.75em}){3-5}\cmidrule(lr{.75em}){6-6}\cmidrule(lr{.75em}){7-9}
        scenario & \multicolumn{1}{c}{$\beta_1$} &  \multicolumn{1}{c}{$k_1$} &  \multicolumn{1}{c}{$r_1$}  & \multicolumn{1}{c}{$\kappa_1$} & \multicolumn{1}{c}{$\beta_2$} & \multicolumn{1}{c}{$k_2$} & \multicolumn{1}{c}{$r_2$} & \multicolumn{1}{c}{$\kappa_2$} \\\midrule
        1 (active-inactive)          &   0.00  &  1.20     &  1.25     &  0.30     &  2.00     &  2.50     &   0.29    &    1.00       \\[0.2em]
        2 (switching preferences)    &  -2.00  &  2.50     &  0.29     &  1.00     &  2.00     &  2.50     &   0.29    &    1.00       \\[0.2em]
        3 (HMM)                      &   0.00  &  1.20     &  1.25     &  0.30     &  0.00     &  2.50     &   0.29    &    1.00       \\
        4 (iSSA, Supp.\ Mat.)       &   2.00  &  2.50     &  0.29     &  1.00     & -        &  --       &   --      &    --         \\
        \bottomrule
    \end{tabular}
    \caption{Overview of the underlying Markov-switching step-selection model parameters for each simulation scenario. The selection coefficients $\beta_i$ for state $i$ ($i=1,2$), describe the habitat selection and belong to the selection function of the model. The movement kernel parameter vector $\boldsymbol{\theta}_i$ includes the shape $k_i$ and rate $r_i$ of gamma distribution for step length and the concentration parameter $\kappa_i$ of the von-Mises distribution for turning angles. Scenario 4 does not include any state-switching and is covered in the Supplementary Material.}
    \label{tab:sim_sen}
\end{table}

In each of the $100$ simulation runs per scenario, we simulated movement paths of length $T=1000$ from the corresponding Markov-switching step-selection model and then applied 2-state MS-iSSAs, 2-state TS-iSSAs and iSSAs to corresponding case-control data sets with $M=20$, $M=100$ and $M=500$ randomly drawn control steps per observed step, respectively. We use different numbers of control locations $M$ to check whether the parameter estimates converge to stable values. For the control steps, we used a uniform distribution for turning angles and a proposal gamma distribution for step length, respectively (Section S2). For model selection purposes, we further estimated the parameters of a $2$-state MS-iSSA with movement but without habitat covariates. This corresponds to a basic movement HMM, but fitted to the same case-control data as the candidate models. All models were implemented in R \citep{RCore22}. For the iSSA, we used the \textit{clogit}-function of the \textit{survival} package \citep{the23}. After parameter estimation, we computed AIC and BIC for model selection \citep{bur02} and the basic p-values of the estimated selection coefficients. For the TS-iSSA and MS-iSSA we further computed the state missclassification rate, i.e.\ the percentage of states that were not correctly classified using the Viterbi algorithm. These metrics were used to evaluate the estimation and classification accuracy of the candidate models, and to assess the performance of standard model selection procedures.


\subsection{Case Study on bank vole interactions}\label{Sec_Case}

To illustrate the use of MS-iSSAs on empirical data, we applied them to movement data of synchronously tracked bank vole individuals (\textit{Myodes glareolus}) as analysed in \citet{schl19}. The data set contains $6$-minute locations of $n=28$ individuals split into $8$ groups, i.e.\ replicates, with $2$ males and $1$-$2$ females each. The individuals within a replicate were synchronously tracked in fenced quadratic outdoor enclosures of $2500\text{m}^2$ for $3-5$ days using collars with small radio telemetry transmitters (1.1 g, BD‐2C, Holohil Systems Ltd., Canada) and a system of automatic receiving units (Sparrow systems, USA). For bank vole individuals tracked under natural conditions, the estimated home range sizes were on average $2029.18\text{m}^2$ with a core area of $549.23\text{m}^2$ \citep{sch19}. Thus, the size of the enclosures allowed the individuals to express their natural movement and space use. Due to daily system maintenance, locations were missing for approximately one hour per day. Otherwise, movement paths were complete. This resulted in $602$–$1,200$ locations per individual split into $3-5$ bursts of around $23$ hours each. 

To study interactions between the bank vole individuals, i.e.\ attraction, avoidance or neutral behaviour towards each other, \citet{schl19} applied SSAs to each individual of each replicate, respectively, using occurrence estimates of the conspecifics as covariates. The occurrence estimate of an individual provides a map of the individual's space use during a certain time window, indicating areas of higher and lower probability of occurrence during that time period. It is estimated from the discrete sample of observed locations through kriging \citep{fle16}. To account for the movement of individuals, occurrence estimates are computed using a rolling time window (here $4$ hours).

The analysis focused on interactions between males and females: Males were expected to mainly show attraction towards females, while females could show any of the three behaviours depending on their reproductive state \citep{schl19}. The authors suggested that the relatively large number of non‐significant interaction coefficients, especially found for male interactions with females, might be caused by unobserved mixtures of different underlying behavioural modes. Bank voles are polyphasic with resting phases of approximately $3\text{h}$ and active phases of approximately $1\text{h}$ following on each other \citep{mir90}. We therefore applied $2$-state MS-iSSAs to the same data to investigate i) if the state-switching model is capable to detect meaningful biological states, and ii) if we find different significant selection, i.e.\ interaction coefficients using the state-switching approach. 

For each individual, we used a $2$-state MS-iSSA with state-dependent gamma distributions for step length, and uniform distribution for turning angle, respectively. Occurrence estimates of each conspecific within the same replicate were used as covariates for the selection part of the model \citep{schl19}. We did not include a resource covariate, as vegetation was sufficiently homogeneous within enclosures. Thus, with $M=500$ available steps per used step, the corresponding selection covariate vector for individual $k$ at time $t$ and locations $\mathbf{x}_{m,t}$, $m=1,\ldots,500$, was given by $\mathbf{Z}_{k,m,t}=(\{O_{-k,m,t}\})$, where $\{O_{-k,m,t}\}$ denotes the set of occurrence estimates of the respective conspecifics withing the same replicate. The corresponding movement covariate vector was $\mathbf{C}_{k,m,t}=(\log(l_{k,m,t}),-l_{k,m,t})$. Parameters were then estimated using a Markov-switching conditional logistic regression with $50$ sets of random starting values to initialise the optimisation (Section \ref{Sec2_est}). For model comparison, we further applied corresponding iSSAs (no state-switching) and  HMMs (no selection) to the same, and $2$-state TS-iSSAs (prior state-classification) to a similar case-control data set for each individual.

\section{Results}\label{Sec_Res}


\subsection{Simulation study}\label{Sec_Res_Sim}

\begin{figure}[!t]
    \includegraphics[width=\textwidth]{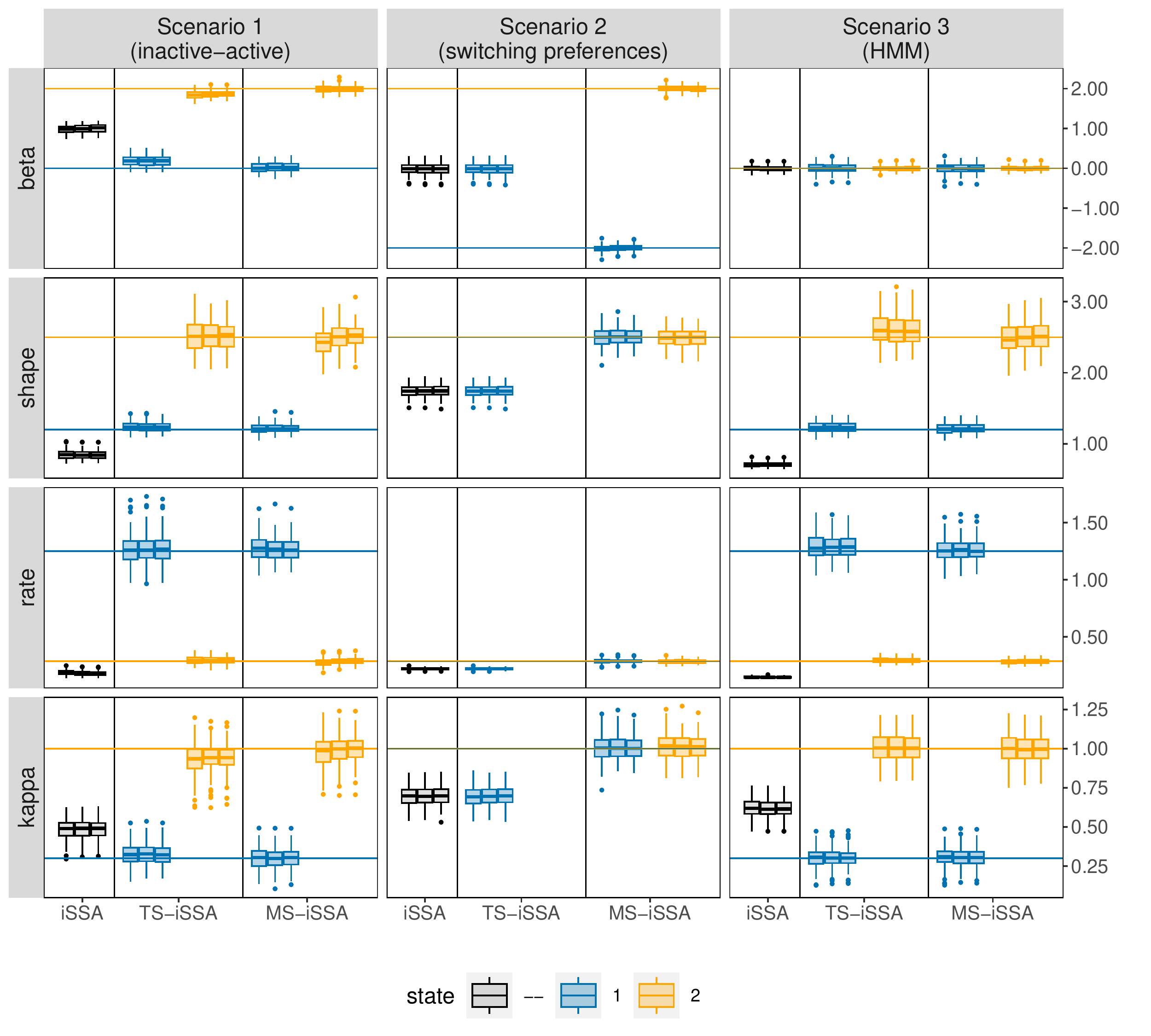}  
\caption{Boxplots of the parameter estimates across the $100$ simulation runs for each applied method, simulation scenario and number of control locations $M$, respectively. The rows refer to the estimated selection coefficient (beta), the shape and rate of the gamma-distribution for step length and the concentration parameter (kappa) of the von Mises distribution for turning angle, respectively. The columns refer to the three different simulation scenarios. For each method (iSSA, TS-iSSA and MS-iSSA) and state (state 1: blue, state 2: orange, no state differentiation: black), the three adjacent boxplots refer the use of $M=20$, $M=100$ and $M=500$ control locations per used location for the parameter estimation. Note that in Scenario 2, the TS-iSSA is naturally not capable to distinguish between two states as both share the same movement kernel. Thus, there are only results for a single state.}
\label{fig:Sim_res}
\end{figure}
Overall, the number of available steps $M$ only slightly affected the estimation results in this simulation exercise, especially the results for $M=100$ and $M=500$ are very similar (Figure \ref{fig:Sim_res}). Thus, the results seem to be stable. The MS-iSSA performed very well across all simulation scenarios and did not produce any evident bias even in the extremer Scenarios 2 (``state-switching preferences'') and 3 (`HMM''; Figure \ref{fig:Sim_res}, Tables S3 and S4). The TS-iSSA was able to detect two suitable states in both scenarios with state-dependent movement kernels (Scenarios 1 and 3), although there was a small but evident bias for some parameters, for example, for the selection coefficients in scenario 1 ($0.18$ in state $1$, $-0.13$ in state $2$ for $M=500$), and for the shape parameter in scenario 3 ($0.12$ in state $2$ for $M=500$). For Scenario 1 (``active-inactive''), this is also reflected in the rather large percentage of significant selection coefficients across the simulation runs in state 1 ($39-40\%$ at a significance level of $\alpha=0.05$, Table \ref{tab:pv}), although the true coefficient is equal to zero. Thus, in contrast to the MS-iSSA, the p-values of the TS-iSSA are not reliable in this active-inactive setting.

\begin{table}[!t]
\centering
\footnotesize
\begin{tabular}{lccccccc}
\toprule
 & & \multicolumn{2}{c}{Scen.\ 1} & \multicolumn{2}{c}{Scen.\ 2} & \multicolumn{2}{c}{Scen.\ 3}\\\cmidrule(lr{.75em}){3-4}\cmidrule(lr{.75em}){5-6}\cmidrule(lr{.75em}){7-8}
\rowcolor{white} model  & no.\ cont.\ & $\beta_1=0$ & $\beta_2=2$ & $\beta_1=-2$ & $\beta_2=2$ & $\beta_1=0$ & $\beta_2=0$ \\\midrule
\multirow{3}{*}{iSSA}       & 20    & 100 & --    & 57 & --     & 16  & -- \\
                            & 100   & 100 & --    & 57 & --     & 17  & -- \\
                            & 500   & 100 & --    & 58 & --     & 17  & -- \\\midrule
\multirow{3}{*}{TS-iSSA}    & 20    & 40  & 100   & 58 & --     & 5   & 6  \\
                            & 100   & 42  & 100   & 57 & --     & 4   & 6  \\
                            & 500   & 39  & 100   & 57 & --     & 4   & 5  \\\midrule
\multirow{3}{*}{MS-iSSA}    & 20    & 2   & 100   & 100 & 100   & 4   & 5  \\
                            & 100   & 3   & 100   & 100 & 100   & 2   & 5  \\
                            & 500   & 1   & 100   & 100 & 100   & 5   & 6  \\
\bottomrule
\end{tabular}
\caption{Percentage of simulation runs in which the selection coefficients are estimated to be significantly different from zero at a significance level of $\alpha=0.05$, for each scenario and fitted model, respectively.}
\label{tab:pv}
\end{table}

\begin{table}[h!t]
\centering
\footnotesize
\begin{tabular}{lcccc}
\toprule
                & \multicolumn{3}{c}{MS-iSSA}  & \\\cmidrule(lr{.75em}){2-4}
    \multicolumn{1}{c}{scenario}    & 20 & 100 & 500 & HMM \\\midrule
    1 (active-inactive) & 4.05 (0.83) & 3.76 (0.82) &  \textbf{3.70} (0.79) & 5.93 (1.47)\\[0.2em]
    2 (switching preferences) & 2.12 (0.53) & 2.00 (0.52) & \textbf{1.94} (0.50) & 49.01 (4.38)\\[0.2em]
    3 (HMM) & 2.49 (0.49) & 2.42 (0.51) & \textbf{2.38} (0.55) & 2.39 (0.53)\\[0.2em]
\bottomrule
\end{tabular}
\caption{Mean missclassification rate with standard deviation in parentheses across the $100$ simulation runs for each scenario and fitted state-switching model, respectively. The missclassification rate is calculated as the percentage of states incorrectly classified using the Viterbi sequence. The lowest missclassification rate for each scenario is highlighted in bold face.}\label{tab:Vit}
\end{table}

The iSSA is by its nature unable to distinguish between the underlying states and thus, did not recover the true underlying parameters in either scenario. Especially in Scenario 2 (``switching preferences''), the iSSA selection coefficients were estimated close to zero and the associated p-values would misleadingly indicate no selection for or against the landscape feature in $42\%-43\%$ of the simulation runs (Table \ref{tab:pv}). Note that the TS-iSSA produced similar results to the iSSA in this scenario, since the inherent HMM classification was not able to distinguish between states that share the same movement kernel, and therefore all steps were classified to belong to the same state.

\begin{table}[h!t]
\centering
\footnotesize
\begin{tabular}[t]{lccccccc}
\toprule
        & & \multicolumn{3}{c}{AIC} & \multicolumn{3}{c}{BIC}\\\cmidrule(lr{.75em}){3-5}\cmidrule(lr{.75em}){6-8}
Scenario  & no.\ cont.\  & iSSA & HMM & MS-iSSA & iSSA & HMM & MS-iSSA    \\\midrule
\multirow{3}{*}{Scen.\ 1}    &   20 & 0 & 0 & \textbf{100} & 0 & 0 & \textbf{100}    \\
                             &   100 & 0 & 0 & \textbf{100} & 0 & 0 & \textbf{100}   \\
                             &   500 & 0 & 0 & \textbf{100} & 0 & 0 & \textbf{100}   \\\midrule
\multirow{3}{*}{Scen.\ 2}   &   20  & 2 & 0 & \textbf{98} & 2 & 0 & \textbf{98}   \\
                            &   100 & 2 & 0 & \textbf{98} & 2 & 0 & \textbf{98}   \\
                            &   500 & 2 & 0 & \textbf{98} & 2 & 0 & \textbf{98}   \\\midrule
\multirow{3}{*}{Scen.\ 3}    &   20  & 0 & \textbf{88} & 12 & 0 & \textbf{100} & 0   \\
                            &   100 & 0 & \textbf{91} & 9 & 0 & \textbf{100} & 0   \\
                            &   500 & 0 & \textbf{86} & 14 & 0 & \textbf{100} & 0   \\
\bottomrule
\end{tabular}
\caption{Percentage of simulation runs in which the three candidate models are selected by either AIC or BIC for each simulation scenario and number of control points used for model fitting, respectively. The cells belonging to the true underlying model are highlighted using bold face.}
\label{tab:ic}
\end{table}

For all three simulation scenarios, the MS-iSSA with $M=500$ available steps achieved the lowest missclassification rate (Table \ref{tab:Vit}). As the data in Scenario 3 were simulated from an HMM, the HMM classification was equally accurate in this scenario. Overall, the MS-iSSA clearly outperformed the other candidate models in its estimation and classification performance in all scenarios.

As the TS-iSSA involves an a-priori HMM classification, it does not provide a proper maximum likelihood value. It is therefore not possible to calculate corresponding AIC or BIC values for model selection. Thus, we only considered iSSAs without state-switching, HMMs without selection (fitted to the same case-control data sets) and MS-iSSAs as candidate models to evaluate information-criteria based model selection in this modelling framework. For Scenario 1 and 2, AIC and BIC performed very well and selected the true underlying model in $100$ and $98\%$ of the simulation runs, respectively (Table \ref{tab:ic}). In Scenario 3 (``HMM''), the AIC tended to select the true HMM model in most of the cases but occasionally selected the more complex MS-iSSA ($9-14\%$ of the runs), while the BIC again selected the correct model in all simulation runs. 

Overall, the simulation runs with lower spatial variation in the landscape variable produced similar results (Section S5.3). However, the lower spatial variation reduces the influence of the habitat selection function on space use. Therefore, the variance in the estimates slightly increased, the HMM missclassification rate decreased in Scenario 1 and the MS-iSSA missclassification rate increased in Scenario 2. In the supplementary simulation scenario without state-switching, the MS-iSSA was able to recovered the true underlying values in state 1, but produced unusable estimates in state 2 (Section S5.2).


\subsection{Case Study on bank vole interactions}\label{Sec_Res_Case}

\begin{figure}[!t]
    \centering
    \includegraphics[width=\textwidth]{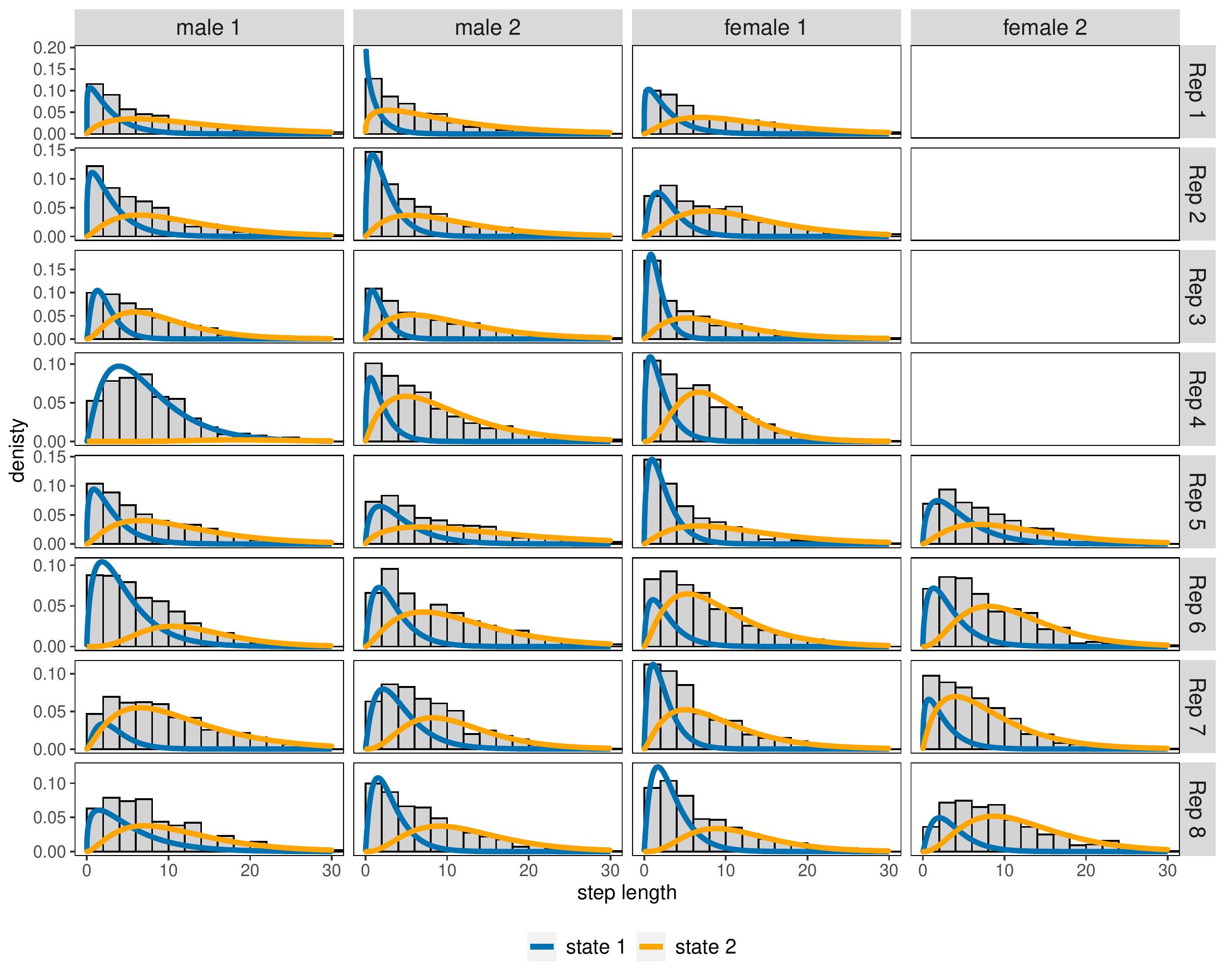}
    \caption{Estimated state-dependent gamma distributions for step length as implied by the fitted $2$-state MS-iSSAs for each individual in replicates $1-8$, respectively. The distributions are weighted by the relative state occupancy frequencies derived from the Viterbi sequence. The gray histograms in the background show the distribution of the observed step lengths.}
    \label{fig:vole_gamma}
\end{figure}

For most bank vole individuals, the MS-iSSA approach could reasonably distinguish between two activity levels. State 1 was always associated to shorter step lengths compared to state 2 which could correspond to a rather inactive behaviour (Figure \ref{fig:vole_gamma}; mean of the estimated gamma distribution for step length ranging from  $1.43$ to $7.38$ in state 1, and from $8.13$ to $20.85$ in state 2, respectively). According to the Viterbi decoded state sequences, the ``less active'' state 1 was occupied between $15.29\%$ and $66.71\%$ of the observed time period (Table S7), except for male 1 in replicate $4$ which spent $96.60\%$ of the time in state 1 according to its decoded state sequence. It is also the individual with the largest estimated mean step length in both states ($7.38$ in state 1, $20.85$ in state 2). Thus, for this male, interpretation must be taken with care. The TS-iSSA provided mostly similar results for the movement kernel and state classification (Figure S10).

\begin{figure}[h!t]
    \centering
    \includegraphics[width=\textwidth]{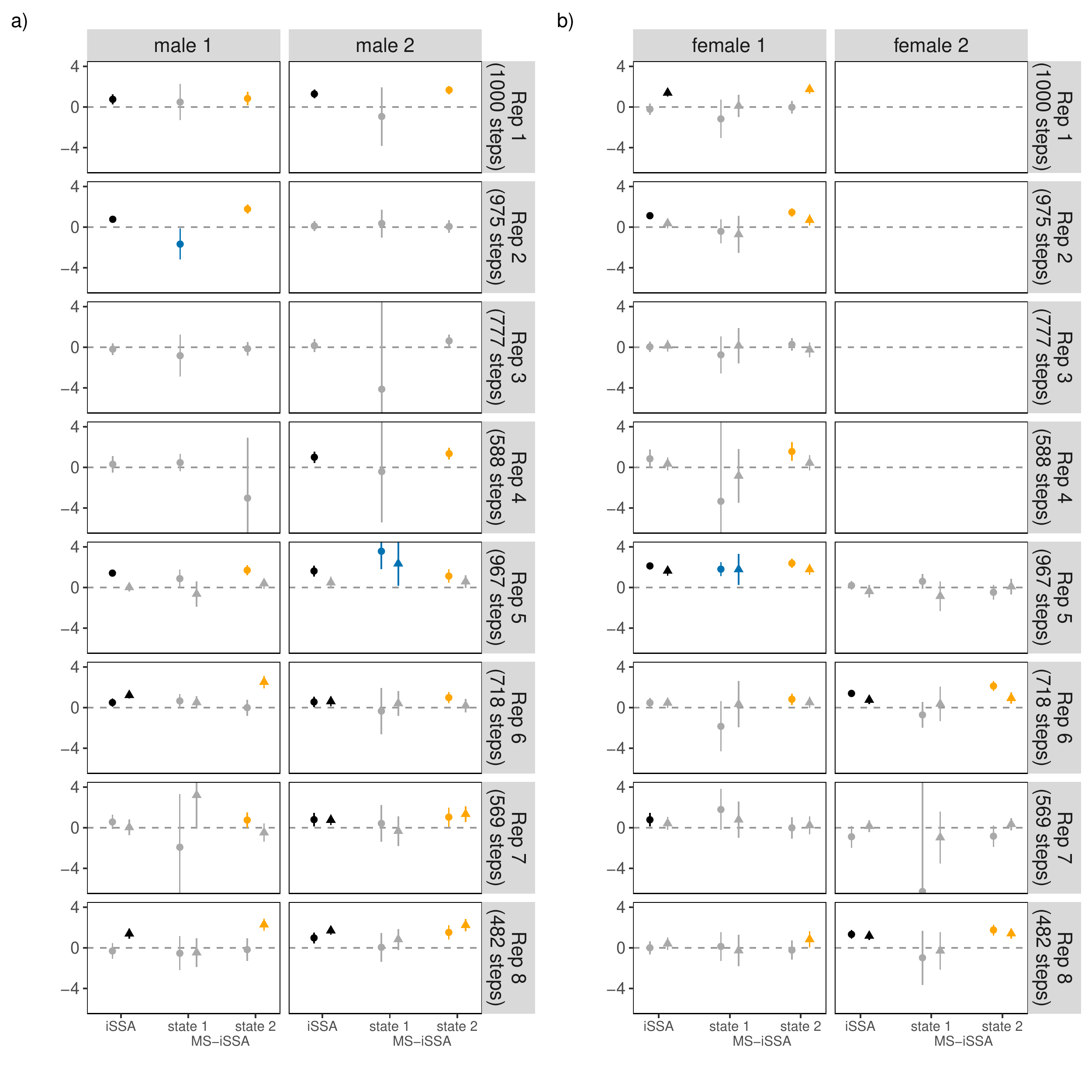}
    \includegraphics[width=0.8\textwidth]{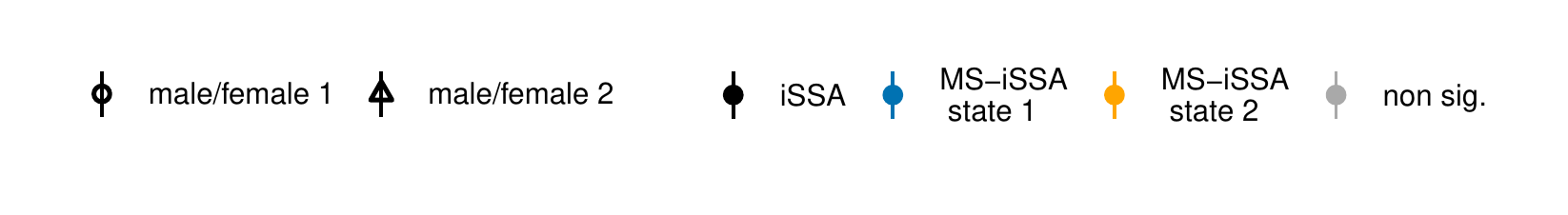}
    \caption{Estimated iSSA and MS-iSSA selection coefficients (solid points/triangles) of interaction behaviour between individuals of opposing sexes within the eight replicates (1–8), including $95\%$ confidence intervals (solid lines). Each replicate consisted of two males (male 1 and male 2) and one or two females (female 1 and female 2) such that each
    individual could respond to up to two opposite‐sex individuals (dot: response to female/male 1, triangle: response to female/male 2 within a replicate). Non‐significant coefficients (p-values below 0.05) are greyed out. The horizontal dashed line indicates zero (i.e. neutral behaviour); positive coefficients indicate attraction, while negative
    coefficients would indicate avoidance.}
    \label{fig:vole_coeff}
\end{figure}

For $21$ of the $28$ bank vole individuals, the MS-iSSA results implied neutral behaviour towards conspecifics in state 1 as all selection coefficients were non-significant ($\alpha=0.05$; Figure \ref{fig:vole_coeff} and Figure S9). This matches well with the interpretation of a less active/inactive state. For two individuals, the results indicated avoidance behaviour in state 1. In state 2 (``active state''), most bank voles showed attraction to at least one bank vole of opposite sex as implied by the positive and significant selection coefficients. However, for four males and four females, the coefficients for occurrence of individuals with opposite sex were non-significant in both states. These are mainly the individuals for which the iSSA also implied neutral behaviour (Figure \ref{fig:vole_coeff}). However, for 3 individuals, i.e.\ male 1 in replicate 7, female 1 in replicate 4, and female 1 in replicate 8, the MS-iSSA indicated attraction towards another individual of opposite sex, while the iSSA indicated neutrality. The opposite is true for female 1 in replicate 7 for which only the iSSA indicated attraction. The selection coefficients for occurrence of individuals with same sex usually implied neutral behaviour in state 1, and neutral or attraction behaviour in state 2 (Figure S9).

Overall, the results of the TS-iSSA are in line with the results of the MS-iSSA (Figures S10, S11 and S12), although the implications are slightly different for nine individuals. Regarding information-criteria based model selection, for most bank voles, AIC and BIC pointed to the Markov-switching step-selection model (Table S8). However, for 10 individuals, including half of the female individuals, BIC selected a simpler model, i.e.\ iSSA or HMM. The selection of HMMs mainly corresponded to cases with many non-significant MS-iSSA selection coefficients. The iSSA was preferred by BIC for male 1 in replicate $4$ and female 2 in replicate 8.


\section{Discussion}\label{Sec_Dis}

In this paper, we discussed the relationship between iSSA without underlying behavioural states, the two-step approach TS-iSSA and the joint approach MS-iSSA and compared them in both a simulation and a case study. Thereby, we highlighted possible consequences of either ignoring underlying behavioural states or using a prior HMM-based state classification to take them into account. This provides important implications for the practical application of fine-scale habitat selection analyses.

Combining ideas of iSSAs and HMMs in a single model, MS-iSSAs build a convenient modelling framework to study state-dependent movement and habitat selection based on animal movement data \citep{nic17,pri22}. This makes a prior state classification unnecessary, which, as demonstrated in the simulation study, could otherwise lead to biased estimates and misleading conclusions (see also \citealp{pri22}). In particular, the MS-iSSA accounts for uncertainties in both the latent state and the observation process which allows for further inference, while the TS-iSSA completely ignores the uncertainties in the state decoding. This renders classical p-values of the TS-iSSA invalid. 

Moreover, the MS-iSSA can detect states associated to same movement but different selection behaviour (Scenario 2 in the simulation study), which is not possible using a prior classification that ignores the selection patterns. While Scenario 2 (``switching preferences'') might cover a rather extreme case, one could imagine, for example, an underlying hungry and a thirsty state where the animal is searching for either food or water, or an attraction and neutrality/avoidance state where the animal is either attracted to another individual or ignoring/avoiding possible social interactions. Even if the movement patterns might not be completely the same across these states, they might largely overlap and therefore lead to problems and high uncertainties in the prior state decoding of the TS-iSSA. This is contrasted with Scenario 3 (``HMM'') of the simulation study which does not include any habitat selection, the states are solely associated with different movement kernels. Here, an HMM-based classification is suitable and the TS-iSSA and MS-iSSA perform equally well. Still, the TS-iSSA does not propagate the uncertainties of the state-decoding.

Our analyses further demonstrates that ignoring underlying behavioural states completely by using standard iSSA can strongly corrupt results on selection behaviour. While theoretically expected, a systematic evaluation and quantification of this effect had been lacking. Our study shows that iSSA tends to average out different selection behaviours in different behavioural states. This can lead to simple over- or underestimation of selection strength, keeping the overall direction of selection (i.e., avoidance or attraction) correct. However, it can also lead to more serious problems when selection behaviour has opposing directions in different states. In this case, we found that selection was estimated to be non-significant, which would lead to a strongly erroneous biological conclusion. This result corroborates the surmise that small effect sizes or non-significant results in step-selection analyses may in fact be due to underlying behavioural state switching \citep{schl19} and more generally the caveat that failure to detect an effect does not imply lack of an affect.

MS-iSSAs have successfully been applied to study habitat selection of bison and zebra in encamped and exploratory states \citep{nic17,pri22}, to detect the onset of mule deer migration and to evaluate the behavioural response of bison on the presence of wolves \citep{pri22}. In our case study, we extend the scope of application to fine-scale interactions of simultaneously tracked bank voles. Here, the 2-state MS-iSSA provided a reasonable separation into a rather inactive state mostly associated with neutral behaviour towards the conspecifics, and an active state often associated with attraction behaviour. However, according to the decoded state sequences, the voles spend more time in the active state as expected ($62.73\%$ of the time on average instead of $25.00\%$). For one male bank vole individual, the state-classification within the MS-iSSA was different. Its second state captured only rare observations with large displacement, while the first state accounted for all other observations. Here, the Viterbi sequence assigned over $96\%$ of the observations to state $1$ and the estimated MS-iSSA showed larger mean step lengths in the estimated state-dependent gamma distributions than for all other individuals. Thus, the second state either captured rare events or outlying observations. This demonstrates that similar care is needed when interpreting the MS-iSSA states as for general HMMs in an unsupervised learning context \citep{mcc20}.

In the active state, we generally expected males to look for females, while females might show different interactions with males depending on the reproductive state \citep{schl19}. For example, females in estrous may actively seek out males to generate mating opportunities away from the nest to lower the risk of infanticide \citep{ecc18}. In contrast, females that are not in estrous state might show avoidance or neutrality toward males. In line with \citet{schl19}, for the male bank voles, we found either attraction or neutral behaviour toward the females in the active state. This was, however, also the case for the female responses to male occurrences. While this might reflect the true individual interaction patterns, it might also be an artefact of measurement errors or the fence around the enclosures that limited the space use. Furthermore, some selection coefficient estimates had rather large confidence intervals possibly associated to the small number of observations for the rather complex model structure. This also prevented the use of a $3$-state model that might have been able to differentiate between pure foraging and social interaction states.

With behavioural states being unobserved, it is usually unclear whether they manifest themselves in a given empirical data set. In both the bank-vole and simulation study, we therefore considered information criteria to select between the candidate models iSSA, HMM and MS-iSSA. Especially the BIC performed well in our simulation study. For the TS-iSSA, such likelihood-based criteria cannot be applied as there is no proper joint maximum log-likelihood value for the state and observation process. This is another drawback of the two-step approach. Besides indicating if the inclusion of states or the inclusion of the selection function are appropriate for a given application, information criteria could also be used to select between MS-iSSAs with different covariate sets or generally to select an appropriate number of meaningful biological states $N$. In the context of HMMs, however, the latter has proven difficult, as information criteria, especially the AIC, tend to select overly complex models with a rather large number of states \citep{cel08,poh17}. We expect this to be the case also for MS-iSSAs. Therefore, besides information criteria, the selection of the number of states should further be based on a close inspection of the fitted models, and involve expert knowledge (``pragmatic order selection'', \citealp{poh17}). This is also highlighted in the supplementary simulation scenario which does not include state-switching. Furthermore, future research could focus on the development of appropriate model checking methods for (Markov-switching) step-selection models.

It is important to note that the resolution of the data in time and space can strongly influence the model results and interpretation. Data sets with different resolutions might reflect different state, movement and selection patterns of an animal \citep{may09,ada19}. For example, an individual can exhibit many behaviours during a long time interval, e.g.\ during $24$ hours, and thus, a coarse time resolution might hinder the model to detect biological states such as resting and foraging or provide only crude state proxies. However, migration modes might be reflected in the data. On the other hand, movement and selection patterns might not directly be expressed in steps at very fine time resolution, e.g.\ based on one location every second \citep{mun21}. Thus, the temporal resolution of the data must match the time scale in which the animal expresses its state, movement and selection patterns of interest. Moreover, if the spatial resolution of a covariate map is too coarse, important habitat features might be overlooked in the analysis \citep{zel17}. Thus, the resolution of the data is a key factor in MS-iSSAs. However, once movement and habitat data are available at a suitable resolution in space and time for a given species and research question at hand, the MS-iSSA approach can flexibly be applied to study fine-scale state-dependent movement and habitat selection. To facilitate its use, the basic MS-iSSA is implemented in the R-package \textit{msissa} available on GitHub \citep{pohle23}.


\section*{Acknowledgements}

The work was supported by the Deutsche Forschungsgemeinschaft (DFG, German Research Foundation, grant no. SCHL 2259/1-1). We thank Sophie Eden, Angela Puschmann and Pauline Lange for help with the bank vole data collection and maintenance of the outdoor enclosures. We declare that there is no conflict of interest.


\section*{Author contributions}
JP, UES and JS conceived the ideas and designed the study. JP implemented the methods with input from UES and JS. JS and JP implemented the R-package. JAE and MD provided the telemetry data and ecological input for the case study. JP led the writing of the manuscript, supported by UES and JS. All authors contributed critically to the drafts and gave final approval for publication.


\section*{Data Availability}
The bank vole data are available from the Dryad Digital Repository: \href{https ://doi.org/10.5061/dryad.rt535m8}{https ://doi.org/10.5061/\\dryad.rt535m8} \citep{schl19}.


\bibliography{bibliography}

\end{spacing}
\end{document}


\begin{spacing}{1.2}
\maketitle

\newpage

\tableofcontents

\newpage

\section{Illustration of state-dependent step-selection densities}
\begin{figure}[ht]
\begin{minipage}{0.65\linewidth}
    \centering
    \subfloat[]{\label{main:a}\includegraphics[width=\textwidth]{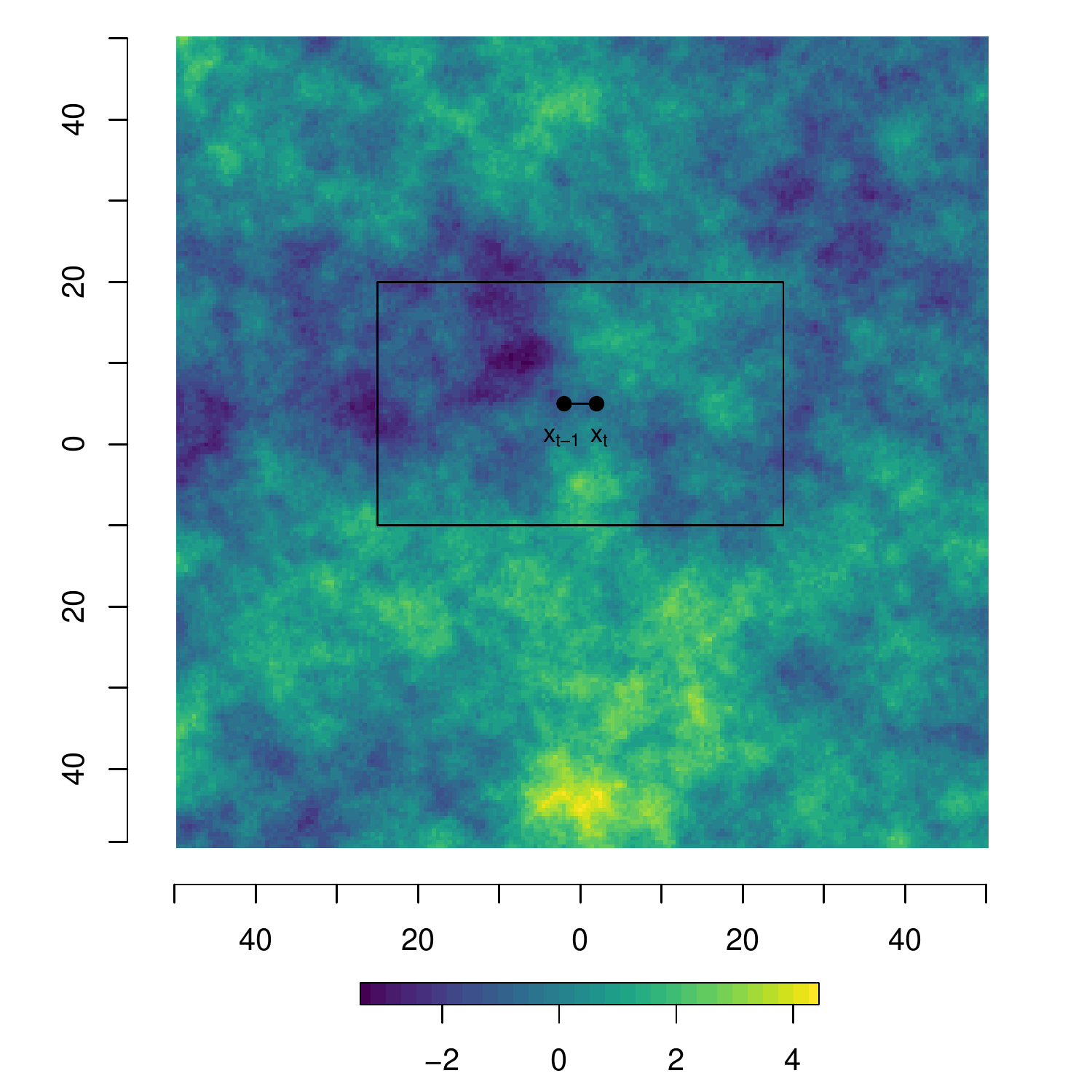}}
\end{minipage}
\begin{minipage}{0.34\linewidth}
    \centering
    \subfloat[]{\label{main:b}\includegraphics[width=\textwidth]{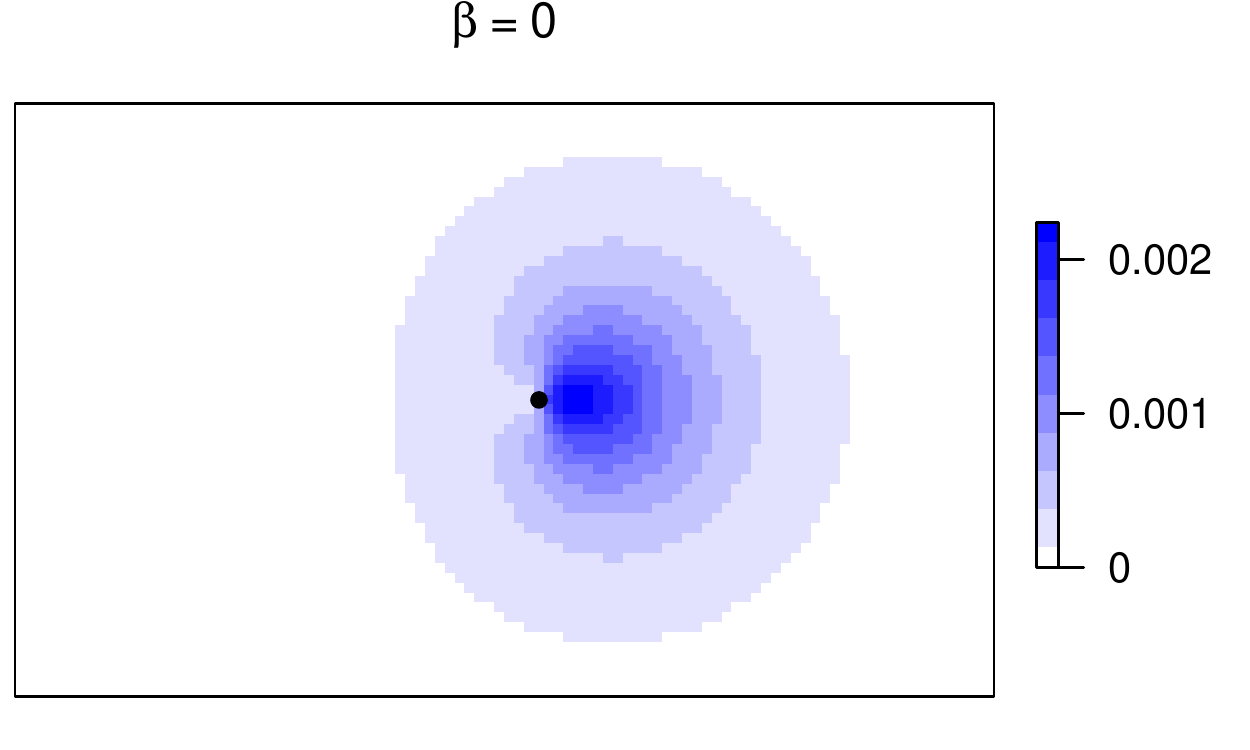}}
        
    \subfloat[]{\label{main:c}\includegraphics[width=\textwidth]{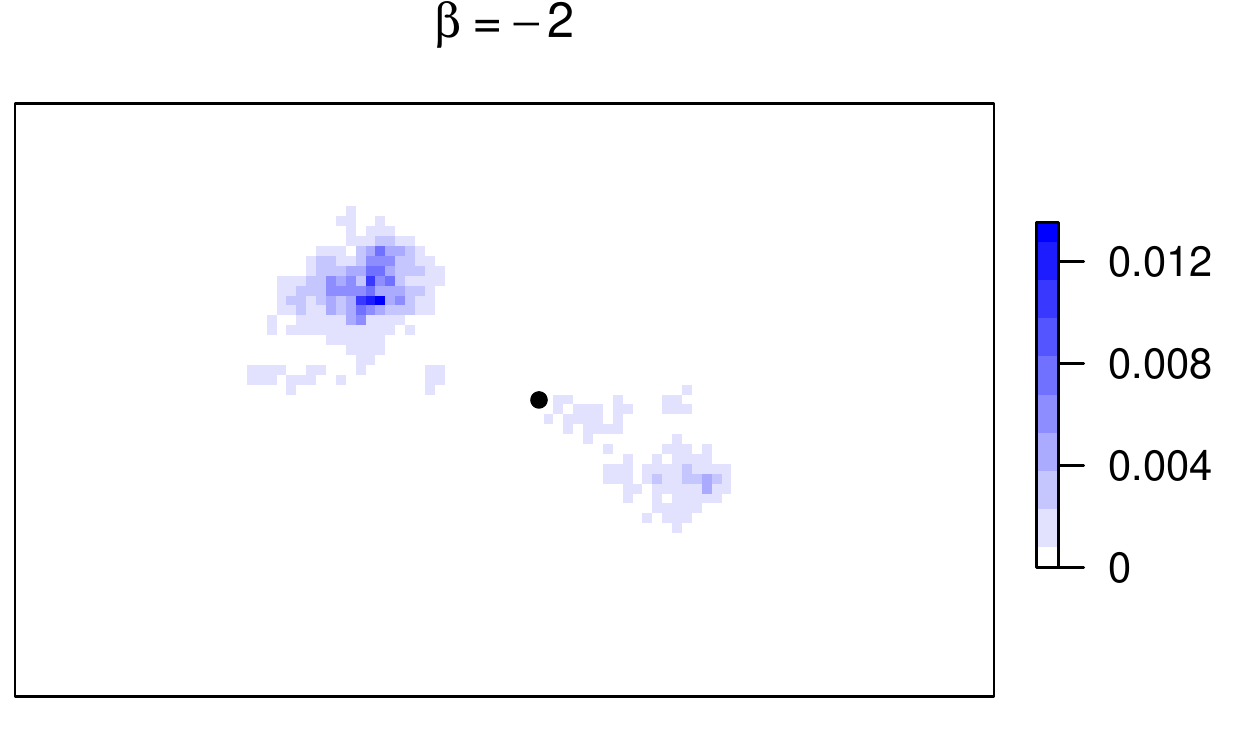}}
    
    \subfloat[]{\label{main:d}\includegraphics[width=\textwidth]{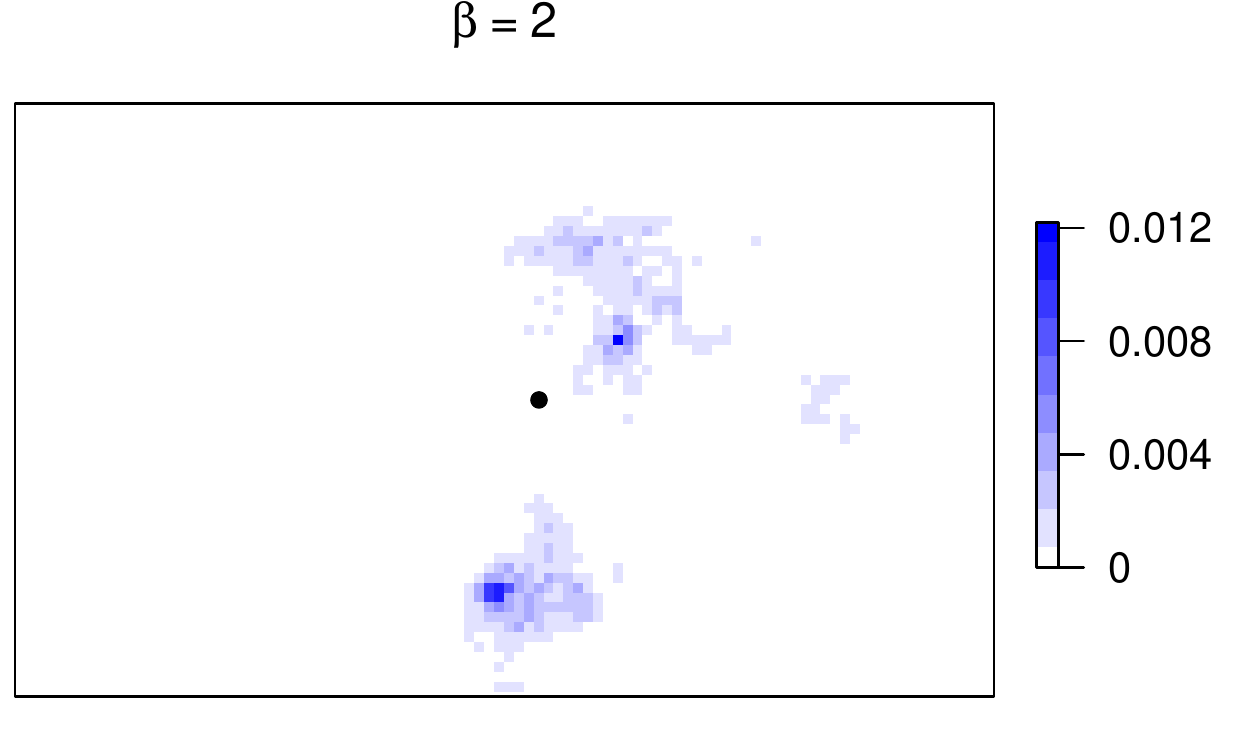}}
\end{minipage}
\caption{Illustration of the state-dependent step selection density $f_i$ in a $2$-state setting as used in Scenario 2 of the simulation study (Section 2.4 main manuscript). Panel a) shows a part of the simulated landscape feature map. The two black dots exemplify the past and current location of the animal. Panel b) shows the movement kernel which consists of a gamma distribution for step length and a von Mises distribution with mean zero for turning angle. In this example, the movement kernel is the same for both states. Panel b) and c) depict the step-selection density for state $1$ ($f_1$) and state $2$ ($f_2$), respectively. In state $1$, the selection coefficient is set to $\beta_1=-2$ which corresponds to avoidance of the landscape feature, while in state $2$, the animal is attracted to the feature ($\beta_2=2$).}
\label{fig:loc_dens}
\end{figure}

\newpage

\section{Movement covariates and sampling procedures}

\subsection{Sampling procedures for the control steps}\label{Sec_samp}

There are different possibilities to sample the control, i.e.\ available steps for the iSSA or MS-iSSA. For iSSAs, usually a procedure related to importance sampling is used which can be described as follows \citep{avg16}: 

\begin{itemize}
    \item[1) ] Proposal distribution:
    \begin{itemize}
        \item[1a)]  Fit the distribution $d_l$ chosen to model step length (e.g.\ gamma distribution) to the observed step lengths $l_{0,t}$ ($t=2,\ldots,T$) to obtain the proposal distribution $d_{l,\text{prop}}$ with corresponding parameter vector $\boldsymbol{\theta}_{l,\text{prop}}$ (e.g.\ containing the shape $k_{\text{prop}}$ and rate $r_{\text{prop}}$ of the fitted gamma distribution).
        \item[1b)] If a von Mises distribution with fixed mean is assumed for turning angle, fit the corresponding von Mises distribution $d_{\alpha}$ to the observed turning angles $\alpha_{0,t}$ ($t=3,\ldots,T$) to obtain the proposal distribution $d_{\alpha,\text{prop}}$ with concentration parameter $\kappa_{\text{prop}}$.
    \end{itemize}
    \item[2) ] Random steps and locations:
    \begin{itemize}
        \item[2a)] For each step starting at $t=2,\ldots,T-1$ , draw $M$ random step length $l_{m,t+1}$ from the proposal distribution $d_{l,{\text{prop}}}$.
        \item[2b)] For each step starting at $t=2,\ldots,T-1$ , further draw $M$ random turning angles $\alpha_{m,t+1}$ either from the proposal distribution $d_{\alpha,\text{prop}}$ if a von Mises distribution is assumed or from a uniform distribution with boundaries $-\pi$ and $\pi$. 
    \item[2c)] Use the $M$ randomly drawn step lengths and turning angles together with the starting point $\mathbf{x}_{0,t}$ to compute the $M$ control locations $\mathbf{x}_{m,t+1}$ and their corresponding movement covariates $\mathbf{C}_{m,t+1}$ ($m=1,\ldots,M$). 
    \end{itemize}
    \item[3) ] Extract the habitat covariates $\mathbf{Z}_{m,t+1}$ for each $\mathbf{x}_{m,t+1}$ from $\mathbf{Z}$ ($m=1,\ldots,M$, $t=2,\ldots,T-1$).
\end{itemize}

Another possibility to sample the control steps is the use of a uniform distribution for both the step length and turning angle of each control step. For the turning angles, values between $-\pi$ and $\pi$ are a natural choice. For step length, however, an upper limit must be chosen. One possibility is to add a constant value, e.g.\ $10$, to the maximum observed step length. Another possibility is to use the $99.9\%$-quantile of the proposal distribution $d_{l,\text{prop}}$ described in step 1a) of the importance sampling procedure described above. Note that the upper step length value must be chosen large enough to cover the main support of the estimated state-dependent step length distributions, otherwise it might lead to an estimation bias \citep{nic17}.

It is possible to combine both sampling approaches. For the simulation study described in Section 2.4 (main manuscript), we drew the random step lengths from a gamma proposal distribution, while the random turning angles were sampled from a uniform distribution (although we assumed a von Mises distribution for turning angles).

Furthermore, it is possible to lay a grid or a mesh either over the complete animals' domain or over a buffer area around the current animal location \citep{schl14,arc23}. For a grid, the mid-points of each grid cell provide the end-point locations for each control step. A fine resolution of the grid improves the estimation accuracy, but might come with high computational costs. Thus, it is important to find a good balance between estimation accuracy and computational costs. This is also true for the number $M$ of randomly drawn control steps. A possible strategy would be to start with a moderate number $M$ (or grid resolution) and increase this number (resolution) until the results stabilise, i.e.\ the estimated parameters hardly change anymore \citep{war10}.

\subsection{Movement parameters and corresponding movement covariates}

Depending on the assumptions made about the step length and turning angle distributions, different movement covariates must be included in analysis \citep{avg16,nic17}. The meaning of the corresponding MS-iSSA movement coefficients is related to the parameters of these distributions, e.g.\ the shape and rate parameters of the gamma distribution, but can slightly differ depending on the sampling procedure used to draw the control locations for the case-control data set. An overview for common step length distributions and the von Mises turning angle distribution is provided in Table \ref{tab:dist_covs}. An overview of how to derive the natural distributional parameters, e.g.\ the shape and rate of the gamma distribution, from the corresponding MS-iSSA movement coefficients is given in Table \ref{tab:dist_back}.
\begin{table}[ht]
    \centering
    \begin{tabular}{llccccc}
    \toprule
                    &                           &                     &                             & \multicolumn{3}{c}{coefficient interpretation}      \\\cmidrule(lr{.75em}){5-7}
                    &                           &                     &                             & ``importance''                                                    & uniform                       & grid \\
    distribution    & parameter                 & covariate           &  coefficient                & sampling                                                      & sampling                      & approach                      \\[0.5em]\toprule
    exponential     & rate $\lambda_i$          & $-l$                & $\theta_{i,-l}$             & $\lambda_i-\lambda_{\text{prop}}$                             & $\lambda_i$                   & $\lambda_i$ (*)                 \\[0.5em]\midrule
    gamma           & shape $k_i$               & $\log(l)$           & $\theta_{i,\log(l)}$        & $k_i-k_{\text{prop}}$                                         & $k_i-1$                       & $k_i-2$                       \\[0.5em]
                    & rate $r_i$                & $-l$                & $\theta_{i,-l}$             & $r_i-r_{\text{prop}}$                                         & $r_i$                         & $r_i$                         \\[0.5em]\midrule
    log-normal      & mean $\mu_i$              & $\log(l)$           & $\theta_{i,\log(l)}$        & $\cfrac{\mu_i}{\sigma_i^2}-\cfrac{\mu_{\text{prop}}}{\sigma_{\text{prop}}^{2}}$  & $\cfrac{\mu}{\sigma^2}-1$       & $\cfrac{\mu}{\sigma^2}-2$     \\[1em]
                    & variance $\sigma_i^2$     & $-\log(l)^2$        & $\theta_{i,-\log(l)^2}$     & $\cfrac{1}{2\sigma_i^2}-\cfrac{1}{2\sigma_{\text{prop}}^2}$    & $\cfrac{1}{2\sigma_i^2}$      & $\cfrac{1}{2\sigma_i^2}$      \\[0.5em]\midrule
    von Mises       & concentration $\kappa_i$  & $\cos(\alpha)$      & $\theta_{i,\cos(\alpha)}$   & $\kappa_i-\kappa_{\text{prop}}$                                & $\kappa_i$                    & $\kappa_i$                    \\
    \bottomrule
    \end{tabular}
    \caption{Example distributions for step length and turning angle. The table displays the distributions' parameters, the associated movement covariates required for the Markov-switching conditional logistic regression in the MS-iSSA, and the meaning of the corresponding movement coefficients when using one of the three sampling procedures to sample the control locations as described in Section \ref{Sec_samp}. (*) For the exponential distribution, $-\log(l)$ must be included as an offset in the grid approach.}
    \label{tab:dist_covs}
\end{table}

\begin{table}[h!t]
    \centering
    \begin{tabular}{llccc}
    \toprule
                    &                           & ``importance''                                                    & uniform                                                   & grid \\
    distribution    & parameter                 & sampling                                                      & sampling                                                  & approach                                              \\[0.5em]\toprule
    exponential     & rate $\lambda_i$          & $\theta_{i,-l}+\lambda_{\text{prop}}$                         & $\theta_{i,-l}$                                           & $\theta_{i,-l}$                                       \\[0.5em]\midrule
    gamma           & shape $k_i$               & $\theta_{i,\log(l)}+k_{\text{prop}}$                          & $\theta_{i,\log(l)}+1$                                    & $\theta_{i,\log(l)}+2$                                \\[0.5em]
                    & rate $r_i$                & $\theta_{i,-l}+r_{\text{prop}}$                               & $\theta_{i,-l}$                                           & $\theta_{i,-l}$                                       \\[0.5em]\midrule
    log-normal      & mean $\mu_i$              & $\cfrac{\theta_{i,\log(l)}+\frac{\mu_{\text{prop}}}{\sigma_{\text{prop}}^{2}}}{2\left(\theta_{i,-\log(l)^2}+\frac{1}{2\sigma^2_{\text{prop}}}\right)}$   & $\cfrac{\theta_{i,\log(l)}+1}{2\theta_{i,-\log(l)^2}}$     & $\cfrac{\theta_{i,\log(l)}+2}{2\theta_{i,-\log(l)^2}}$  \\[2em]
                    & variance $\sigma_i^2$     & $\cfrac{1}{2}\left(\theta_{i,-\log(l)^2}+\frac{1}{2\sigma^2_{\text{prop}}}\right)^{-1}$   & $\cfrac{1}{2\theta_{i,-\log(l)^2}}$        & $\cfrac{1}{2\theta_{i,-\log(l)^2}}$                    \\[0.5em]\midrule
    von Mises       & concentration $\kappa_i$  & $\theta_{i,\cos(\alpha)}+\kappa_{\text{prop}}$                & $\theta_{i,\cos(\alpha)}$                                 & $\theta_{i,\cos(\alpha)}$                             \\
    \bottomrule
    \end{tabular}
    \caption{Overview of deriving the natural step length and turning angle parameters from the MS-iSSA movement coefficients for the three control step sampling procedures described in Section \ref{Sec_samp}.}
    \label{tab:dist_back}
\end{table}

\clearpage

\section{Initial values for the maximum likelihood estimation}

The MS-iSSA requires starting values, i.e.\ first guesses, for each model parameter to initialise the parameter estimation. For each state $i=1,\ldots,N$ the model comprises:
\begin{enumerate}
    \item the state transition probabilities $\gamma_{ij}=\Pr(S_{t}=j\mid S_{t-1}=i)$ with $j=1,\ldots,N$:
    \begin{itemize}
        \item probabilities for the underlying state sequence to switch from state $i$ to state $j=1,\ldots,N$
        \item values lie between zero and one, i.e.\ $0 \leq \gamma_{ij} \leq 1$
        \item values must sum to one, i.e.\ $\sum_{j=1}^{N} \gamma_{ij} = 1$
        \item usually, a certain persistence in the states is assumed, such that the probability to remain in the current state is rather high, e.g.\ values between $0.8$ and $0.95$
    \end{itemize}
    \item the initial state probability $\delta_{i}=\Pr(S_{t_{0}}=i)$:
    \begin{itemize}
        \item probability for the state sequence to start in state $i$
        \item values lie between zero and one, i.e.\ $0 \leq \delta_{i} \leq 1$
        \item if stationarity is assumed for the Markov chain, the initial probabilities are not estimated but computed from the model's transition probability matrix
    \end{itemize}
    \item the state-dependent movement parameter vector $\boldsymbol{\theta}_{i}$:
    \begin{itemize}
        \item contains the parameters of the movement kernel
        \item depends on the distributions assumed for step length and turning angle (see Table \ref{tab:dist_covs})
        \item for gamma distributed step length and von-Mises distributed turning angles, it comprises the shape $k_i$ and rate $r_i$ of the gamma distribution and the concentration parameter $\kappa_i$ of the von Mises distribution; all values must be greater than zero
    \end{itemize}
    \item the state-dependent selection coefficient vector $\boldsymbol{\beta}_i$
    \begin{itemize}
        \item contains the parameters for the selection function 
        \item comprises the selection coefficients for each habitat covariate 
    \end{itemize}
\end{enumerate}

The log-likelihood (LL) function of the Markov-switching step selection model is usually multi-modal. Thus, it is possible that the optimisation procedure ends up in a local maximum instead of the global maximum of the log-likelihood (MLL) required for the maximum likelihood estimates (MLEs). As the starting values build the starting point for the search of the function's maximum, they can have a large impact on the output of the optimisation algorithm. Therefore, its is necessary to run the MS-iSSA with different sets of (random) starting values to increase the chance of finding the MLEs. Among the set of MS-iSSAs fitted with different starting values, the one with the largest corresponding MLL value provides the final results. However, to ensure numerical stability and convergence, it is necessary to check if different sets of initial values have lead to the same MLL value and MLEs. This can be done, for example, by plotting the MLL values found across the different initial values and by comparing the estimates of MS-iSSAs with similar MLL values. 

It can be inefficient to choose the starting values completely at random which can result in implausible values and in turn lead to poor model results (e.g.\ singular models). Therefore, it is usually useful to restrict the initial parameter values to plausible ranges and to ensure that the initial parameter values differ between the $N$ states. An inspection of the observed step length and turning angle observations, e.g.\ plotting the histograms and the raw data, might be helpful to find such plausible ranges.

In the following, we describe a possible procedure for selecting the starting values, which can serve as a starting point and can be adapted depending on the data and model at hand:
\begin{enumerate}
    \item Choose large random values for the probabilities to remain in a given state ($\gamma_{ii}$), e.g.\ draws from a uniform distribution between $0.80$ and $0.95$ (\textit{runif}($N,0.8,0.95$) in R). 
    
    For the remaining transition probabilities, $\gamma_{ij}=\cfrac{1-\gamma_{ii}}{N-1}$ can be used for $i \neq j$ . Otherwise, when drawing random starting values for the off-diagonals of the transition probability matrix, make sure that transition probabilities from state $i$ to any other state sum to 1.
    \item If no stationarity is assumed for the underlying latent Markov chain, the starting values for the initial state distribution can be fixed to $\delta_i=\cfrac{1}{N}$ or randomly drawn from a uniform distribution with values between $0$ and $1$. For the latter, the values must then be normalised with $\cfrac{\delta_i}{\sum_{i=1}^{N}\delta_i}$.
    \item For the movement kernel, initial parameters for the step length distribution and possibly for the turning angle distribution are required.
    \begin{itemize}
        \item Step length distribution: the quantiles of the step length observations can be used to draw random mean step length values $\mu_i$ (\textit{quantile}-function in R). To avoid a large overlap between the mean values across the states, $N+1$ step length quantiles $q_1,\ldots,q_{N+1}$ can be considered for $N$ states, for example with $q_1$ being the $0.1$-quantile and $q_{N+1}$ being the $0.9$-quantile. The mean step length value $\mu_i$ for state $i$ is then drawn from a uniform-distribution with values between $q_{i}$ and $q_{i+1}$. The corresponding standard deviation $\sigma_i$ can be drawn from a uniform distribution, for example, with values between a quarter and twice the mean value, i.e. $\cfrac{1}{4}\mu_i$ and $2\mu_i$. Based on these state-dependent mean values and standard deviations, the corresponding step length distribution parameters can be derived. For a gamma distribution, this is the shape with $k_i=\cfrac{\mu_i^2}{\sigma_i^2}$ and rate with $r_i=\cfrac{\mu_i}{\sigma_i^2}$.
        \item Turning angles: The concentration parameter of the von Mises distribution can be randomly drawn from a uniform distribution with values between $0.2$ and $2$.
    \end{itemize}
    \item For the selection coefficients, draw random values from a set of plausible values, e.g.\ $\{-2.0,-1.0,$ $-0.5,0.0,0.5,1.0,2.0\}$ (using the \textit{sample}-function in R).
    \item repeat the procedure several, e.g.\ $50$ times.
\end{enumerate}

As a starting point, our R-package \textit{msissa} provides a function named \textit{generate\_starting\_values} to draw random starting values for the MS-iSSA.

\clearpage

\section{Supplementary material for the simulation study}

\subsection{Simulated landscape feature}

\begin{figure}[ht]
    \centering
    \includegraphics[width=\textwidth]{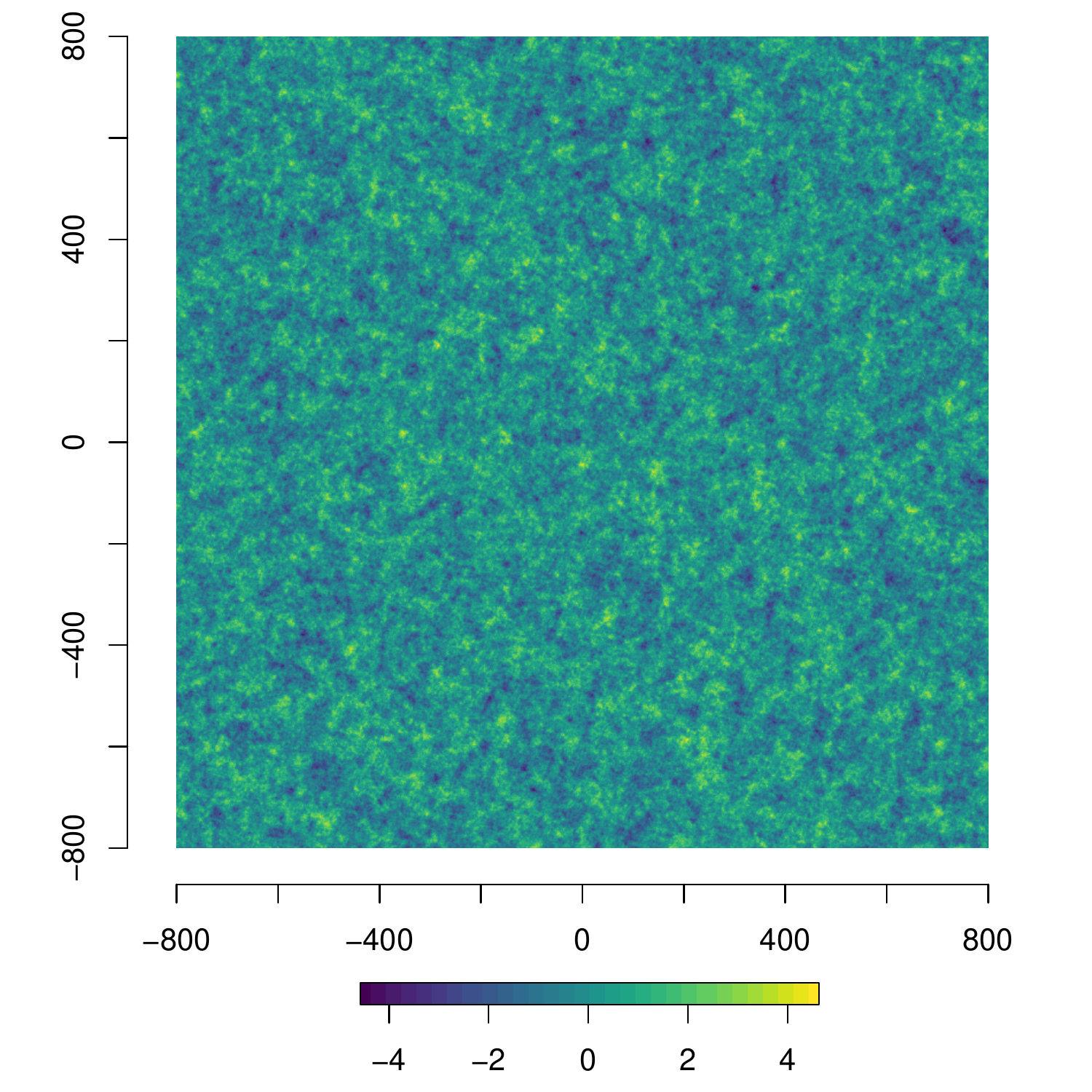}
    \caption{Simulated landscape feature map used to derive the habitat covariates in the simulation study described in Section 2.4 (main manuscript).}
    \label{fig:land_sim}
\end{figure}

\newpage

\subsection{Bias}

\begin{table}[ht]
\centering
\begin{tabular}[t]{lllcccccccc}
\toprule
& & & \multicolumn{2}{c}{select.\ coeff.\ } & \multicolumn{2}{c}{shape} & \multicolumn{2}{c}{rate} & \multicolumn{2}{c}{conc.\ } \\\cmidrule(lr{.75em}){4-5}\cmidrule(lr{.75em}){6-7}\cmidrule(lr{.75em}){8-9}\cmidrule(lr{.75em}){10-11}
 scenario & method & no.\ cont.\ & $\beta_1$ & $\beta_2$ & $k_1$ & $k_2$ & $r_1$ & $r_2$ & $\kappa_1$ & $\kappa_2$   \\
\midrule
\multirow{9}{*}{1} & \multirow{3}{*}{iSSA} & 20 & 0.98 & -1.02 & -0.35 & -1.65 & -1.06 & -0.10 & 0.18 & -0.52\\
            &       & 100 & 1.00 & -1.00 & -0.35 & -1.65 & -1.07 & -0.10 & 0.19 & -0.51\\
            &       & 500 & 1.01 & -0.99 & -0.35 & -1.65 & -1.07 & -0.11 & 0.19 & -0.51\\\cmidrule{2-11}
            & \multirow{3}{*}{TS-iSSA} & 20 & 0.19 & -0.16 & 0.04 & 0.01 & 0.02 & 0.01 & 0.02 & -0.07\\
            &       & 100 & 0.18 & -0.14 & 0.04 & 0.02 & 0.02 & 0.01 & 0.03 & -0.06\\
            &       & 500 & 0.18 & -0.13 & 0.04 & 0.02 & 0.02 & 0.01 & 0.02 & -0.06\\\cmidrule{2-11}
            & \multirow{3}{*}{MS-iSSA} & 20 & 0.01 & -0.01 & 0.01 & -0.07 & 0.03 & -0.01 & 0.00 & -0.02\\
            &       & 100 & 0.02 & -0.01 & 0.02 & 0.00 & 0.02 & 0.00 & 0.00 & 0.00\\
            &       & 500 & 0.02 & -0.01 & 0.02 & 0.03 & 0.02 & 0.00 & 0.00 & 0.00\\\midrule\midrule
\multirow{9}{*}{2} & \multirow{3}{*}{iSSA} & 20 & 1.99 & -2.01 & -0.76 & -0.76 & -0.06 & -0.06 & -0.30 & -0.30\\
            &       & 100 & 1.99 & -2.01 & -0.76 & -0.76 & -0.06 & -0.06 & -0.30 & -0.30\\
            &       & 500 & 1.99 & -2.01 & -0.76 & -0.76 & -0.06 & -0.06 & -0.30 & -0.30\\\cmidrule{2-11}
            & \multirow{3}{*}{TS-iSSA} & 20 & 1.96 & -1.97 & -0.76 & -0.76 & -0.06 & -0.06 & -0.30 & -0.30\\
            &       & 100 & 1.96 & -1.97 & -0.76 & -0.75 & -0.06 & -0.06 & -0.30 & -0.30\\
            &       & 500 & 1.96 & -1.97 & -0.76 & -0.75 & -0.06 & -0.06 & -0.30 & -0.30\\\cmidrule{2-11}
            & \multirow{3}{*}{MS-iSSA} & 20 & -0.02 & 0.01 & -0.01 & -0.03 & 0.00 & 0.00 & 0.00 & 0.00\\
            &       & 100 & -0.01 & 0.00 & -0.01 & -0.03 & 0.00 & 0.00 & 0.00 & 0.00\\
            &       & 500 & 0.00 & -0.01 & -0.01 & -0.03 & 0.00 & 0.00 & -0.02 & -0.02\\\midrule\midrule
\multirow{9}{*}{3} & \multirow{3}{*}{iSSA} & 20 & -0.01 & -0.01 & -0.50 & -1.80 & -1.10 & -0.14 & 0.32 & -0.38\\
            &       & 100 & -0.01 & -0.01 & -0.49 & -1.79 & -1.10 & -0.14 & 0.32 & -0.38\\
            &       & 500 & -0.01 & -0.01 & -0.49 & -1.79 & -1.10 & -0.14 & 0.32 & -0.38\\\cmidrule{2-11}
            & \multirow{3}{*}{TS-iSSA} & 20 & 0.00 & 0.00 & 0.03 & 0.12 & 0.04 & 0.01 & 0.00 & 0.01\\
            &       & 100 & 0.00 & 0.00 & 0.03 & 0.12 & 0.04 & 0.01 & 0.00 & 0.01\\
            &       & 500 & 0.00 & 0.00 & 0.03 & 0.12 & 0.04 & 0.01 & 0.00 & 0.01\\\cmidrule{2-11}
            & \multirow{3}{*}{MS-iSSA} & 20 & 0.00 & 0.00 & 0.01 & -0.02 & 0.01 & 0.00 & 0.00 & 0.00\\
            &       & 100 & 0.00 & 0.00 & 0.01 & 0.01 & 0.01 & 0.00 & 0.00 & 0.00\\
            &       & 500 & 0.00 & 0.00 & 0.02 & 0.02 & 0.01 & 0.00 & 0.00 & 0.00\\
\bottomrule
\end{tabular}
\caption{Estimation bias for the model parameters for each simulation scenario and applied method. The bias is calculated as the mean difference between the parameter estimate and the true parameter value across the $100$ simulation runs.}
\end{table}

\newpage

\subsection{Root mean squared error}

\begin{table}[ht]
\centering
\begin{tabular}[t]{llccccccccc}
\toprule
& & & \multicolumn{2}{c}{select.\ coeff.\ } & \multicolumn{2}{c}{shape} & \multicolumn{2}{c}{rate} & \multicolumn{2}{c}{conc.\ } \\\cmidrule(lr{.75em}){4-5}\cmidrule(lr{.75em}){6-7}\cmidrule(lr{.75em}){8-9}\cmidrule(lr{.75em}){10-11}
 scenario & model & no.\ cont.\ & $\beta_1$ & $\beta_2$ & $k_1$ & $k_2$ & $r_1$ & $r_2$ & $\kappa_1$ & $\kappa_2$   \\
\midrule
\multirow{9}{*}{1} & \multirow{3}{*}{iSSA} & 20 & 0.98 & 1.03 & 0.36 & 1.65 & 1.06 & 0.10 & 0.19 & 0.52\\
            &       & 100 & 1.01 & 1.01 & 0.36 & 1.65 & 1.07 & 0.11 & 0.20 & 0.52\\
            &       & 500 & 1.01 & 1.00 & 0.36 & 1.66 & 1.07 & 0.11 & 0.20 & 0.52\\\cmidrule{2-11}
            & \multirow{3}{*}{TS-iSSA} & 20 & 0.23 & 0.18 & 0.08 & 0.23 & 0.13 & 0.04 & 0.08 & 0.13\\
            &       & 100 & 0.23 & 0.16 & 0.08 & 0.21 & 0.13 & 0.04 & 0.07 & 0.11\\
            &       & 500 & 0.22 & 0.15 & 0.08 & 0.21 & 0.13 & 0.03 & 0.07 & 0.11\\\cmidrule{2-11}
            & \multirow{3}{*}{MS-iSSA} & 20 & 0.12 & 0.10 & 0.07 & 0.20 & 0.12 & 0.04 & 0.07 & 0.11\\
            &       & 100 & 0.12 & 0.08 & 0.07 & 0.17 & 0.11 & 0.03 & 0.07 & 0.09\\
            &       & 500 & 0.12 & 0.07 & 0.07 & 0.17 & 0.10 & 0.03 & 0.07 & 0.09\\\midrule\midrule
\multirow{9}{*}{2} & \multirow{3}{*}{iSSA} & 20 & 1.99 & 2.01 & 0.77 & 0.77 & 0.07 & 0.07 & 0.31 & 0.31\\
            &       & 100 & 1.99 & 2.02 & 0.76 & 0.76 & 0.07 & 0.07 & 0.31 & 0.31\\
            &       & 500 & 1.99 & 2.02 & 0.76 & 0.76 & 0.07 & 0.07 & 0.31 & 0.31\\\cmidrule{2-11}
            & \multirow{3}{*}{TS-iSSA} & 20 & 1.99 & 2.01 & 0.77 & 0.77 & 0.07 & 0.07 & 0.31 & 0.31\\
            &       & 100 & 1.99 & 2.02 & 0.76 & 0.76 & 0.07 & 0.07 & 0.31 & 0.31\\
            &       & 500 & 1.99 & 2.02 & 0.76 & 0.76 & 0.07 & 0.07 & 0.31 & 0.31\\\cmidrule{2-11}
            & \multirow{3}{*}{MS-iSSA} & 20 & 0.09 & 0.09 & 0.18 & 0.17 & 0.02 & 0.02 & 0.10 & 0.10\\
            &       & 100 & 0.08 & 0.08 & 0.17 & 0.17 & 0.02 & 0.02 & 0.09 & 0.09\\
            &       & 500 & 0.08 & 0.08 & 0.16 & 0.17 & 0.02 & 0.02 & 0.16 & 0.17\\\midrule\midrule
\multirow{9}{*}{3} & \multirow{3}{*}{iSSA} & 20 & 0.07 & 0.07 & 0.50 & 1.80 & 1.10 & 0.14 & 0.33 & 0.38\\
            &       & 100 & 0.07 & 0.07 & 0.50 & 1.80 & 1.10 & 0.14 & 0.33 & 0.38\\
            &       & 500 & 0.07 & 0.07 & 0.50 & 1.80 & 1.10 & 0.14 & 0.33 & 0.38\\\cmidrule{2-11}
            & \multirow{3}{*}{TS-iSSA} & 20 & 0.12 & 0.06 & 0.08 & 0.26 & 0.11 & 0.03 & 0.07 & 0.09\\
            &       & 100 & 0.12 & 0.06 & 0.08 & 0.26 & 0.11 & 0.03 & 0.06 & 0.09\\
            &       & 500 & 0.12 & 0.06 & 0.08 & 0.26 & 0.11 & 0.03 & 0.06 & 0.09\\\cmidrule{2-11}
            & \multirow{3}{*}{MS-iSSA} & 20 & 0.13 & 0.07 & 0.07 & 0.22 & 0.10 & 0.02 & 0.07 & 0.10\\
            &       & 100 & 0.12 & 0.06 & 0.07 & 0.21 & 0.10 & 0.02 & 0.07 & 0.09\\
            &       & 500 & 0.12 & 0.06 & 0.07 & 0.21 & 0.10 & 0.02 & 0.07 & 0.09\\
\bottomrule
\end{tabular}
\caption{Root mean squared error of the estimated parameters calculated across the $100$ simulation runs for each fitted model and simulation scenario, respectively. The root mean squared error is calculated as the square root of the average squared difference between the parameter estimate and the true parameter value across the $100$ simulation runs.}
\label{tab_supp:RMSE2}
\end{table}

\clearpage

\section{Additional simulation runs}

\subsection{Uniform sampling of the control locations}

\begin{figure}[ht]
    \centering
    \includegraphics[width=\textwidth]{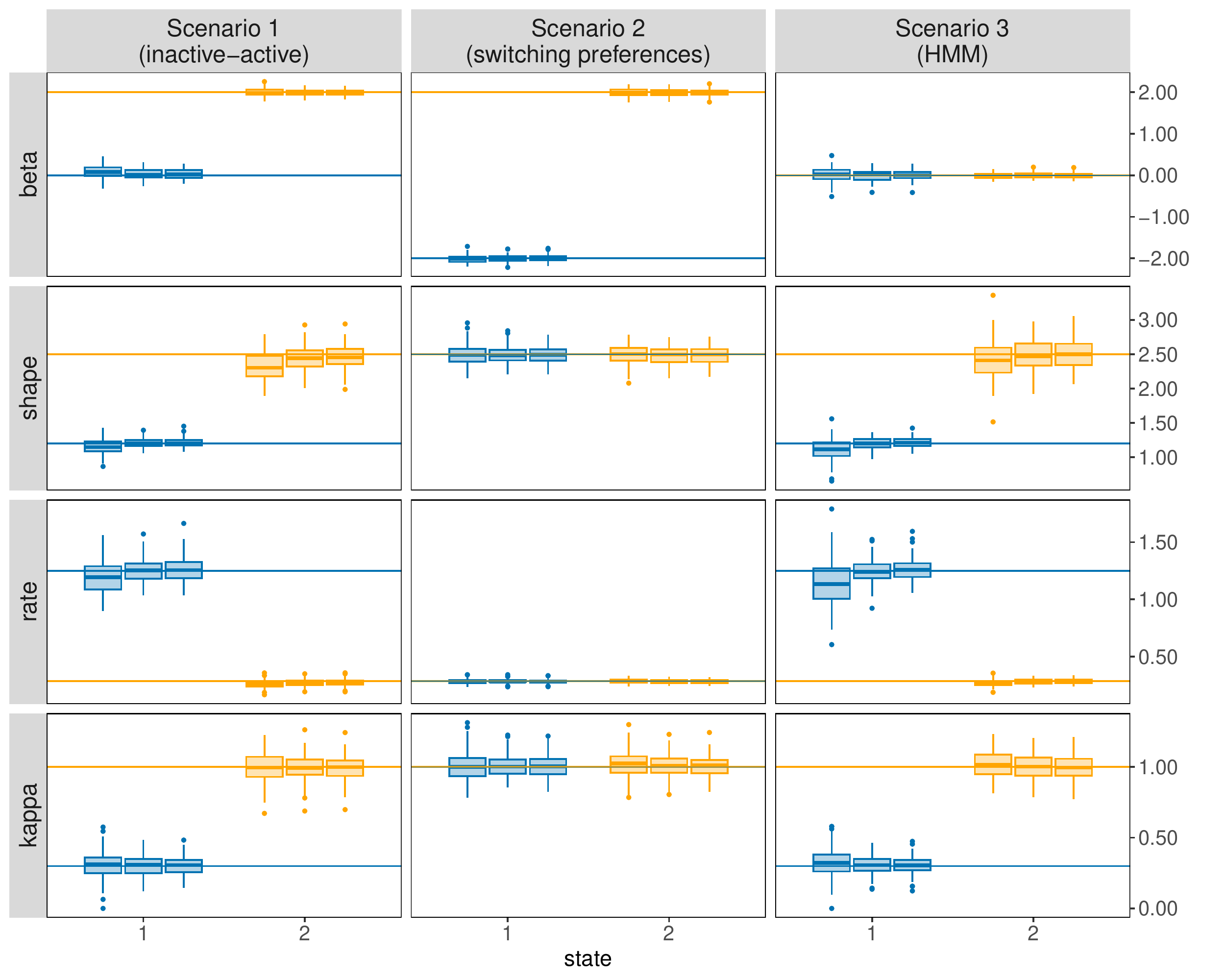}
    \caption{Uniform sampling of the control locations: Boxplots of the MS-iSSA parameter estimates across the $100$ simulation runs for each simulation scenario and number of control locations $M$, respectively. The control locations for the case-control locations were derived from uniformly sampled step lengths and turning angles. Here, for the upper step length limit, $10$ was added to the maximum observed step length. The rows refer to the estimated selection coefficient (beta), the shape and rate of the gamma distribution for step length and the concentration parameter (kappa) of the von Mises distribution for turning angle, respectively. The columns refer to the three different simulation scenarios. The colours indicate the different states (state 1: blue, state 2: orange), the three adjacent boxplots refer the use of $M=20$, $M=100$ and $M=500$ control locations per used location for the parameter estimation.}
    \label{fig:unif}
\end{figure}

We re-ran the simulation study described in Section 2.4 (main manuscript) with uniformly sampled random steps for the MS-iSSA. The results are comparable to the results described in the main manuscript (Figure \ref{fig:unif}). However, with uniform sampling, the results seem to be more sensitive to the number $M$ of control steps used for each observed step. For example, with $M=20$ control locations, the rate parameters in Scenario 1 and 2 are slightly biased. Thus, $M=20$ control locations might be insufficient here.

\clearpage

\subsection{Scenario without state-switching}

To check the robustness of the MS-iSSA for data without underlying behavioral states, we ran a fourth simulation scenario with data generated from a standard iSSA. For the movement-kernel, we used gamma distributed step length with shape $k=2.5$ and rate $r=0.29$ and von-Mises distributed turning angles with mean zero and concentration $\kappa=1$. The selection coefficient was set to $\beta=2$. A corresponding $2$-state MS-iSSA would need to share the same movement and selection coefficients across both states (see Section 2.1, main manuscript). However, as state $1$ and $2$ do not differ, a switch from one state to the other can occur arbitrarily. Therefore, the transition probability matrix is not identified in this scenario and could take any form. This could lead to numerical problems and instabilities in the MS-iSSA parameter estimation. 

\begin{figure}[!b]
    \centering
    \includegraphics[width=\textwidth]{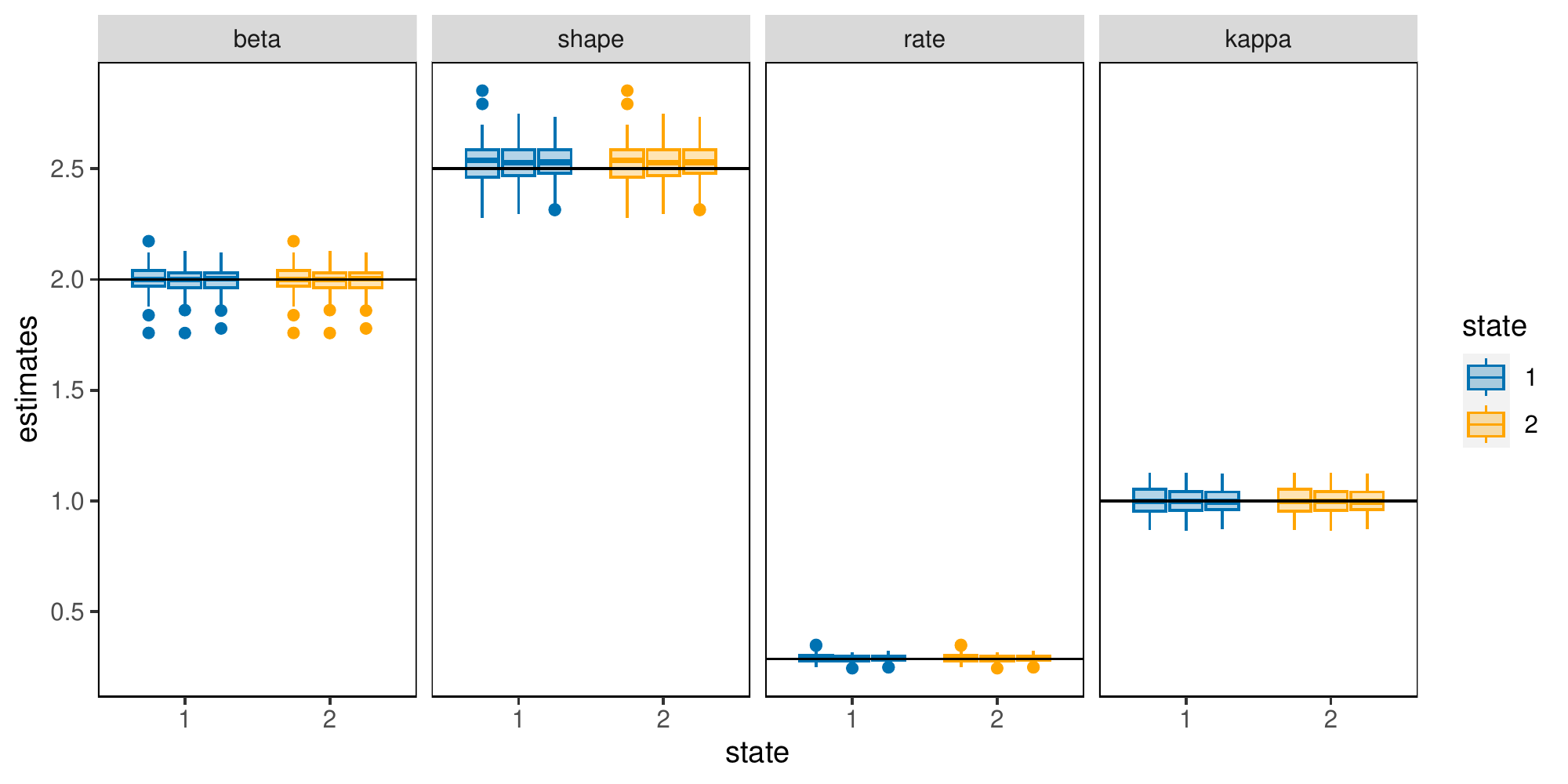}
    \caption{Boxplots of the MS-iSSA parameter estimates across the $100$ simulation runs for the iSSA-scenario for each number of control locations $M$, respectively. Here, the true values were used as starting values for the Markov-switching conditional logistic regression. The figure columns refer to the estimated selection coefficient (beta), the shape and rate of the gamma distribution for step length and the concentration parameter (kappa) of the von Mises distribution for turning angle, respectively. The colours indicate the different states (state 1: blue, state 2: orange), the three adjacent boxplots refer the use of $M=20$, $M=100$ and $M=500$ control locations per used location for the parameter estimation.}
    \label{fig:ssf0}
\end{figure}

\begin{figure}[ht]
    \centering
    \includegraphics[width=\textwidth]{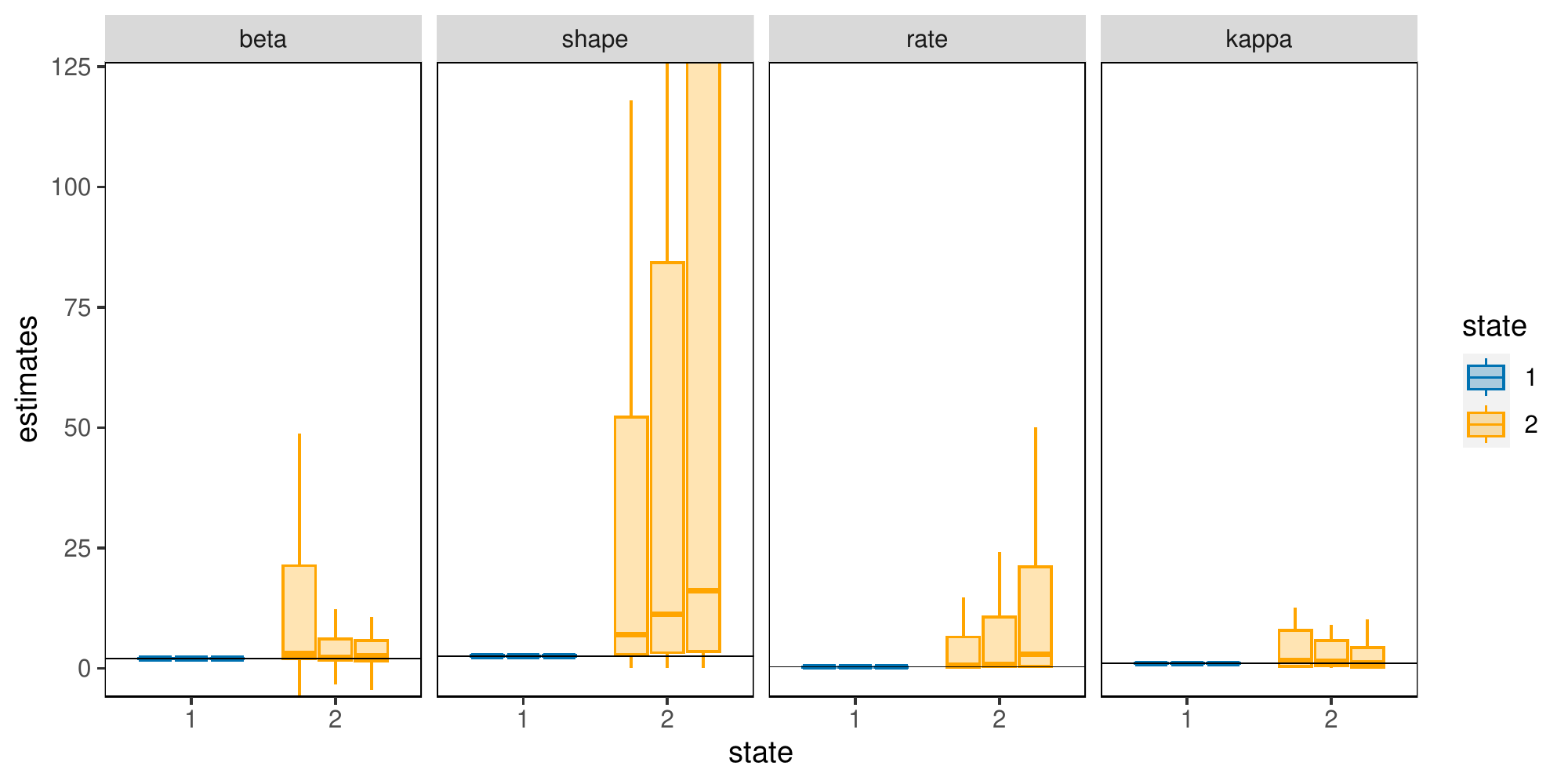}
    \caption{Boxplots of the MS-iSSA parameter estimates across the $100$ simulation runs for the iSSA-scenario for each number of control locations $M$, respectively. Here, $50$ random starting values were used as starting values for the Markov-switching conditional logistic regression. The figure columns refer to the estimated selection coefficient (beta), the shape and rate of the gamma distribution for step length and the concentration parameter (kappa) of the von Mises distribution for turning angle, respectively. The colours indicate the different states (state 1: blue, state 2: orange), the three adjacent boxplots refer the use of $M=20$, $M=100$ and $M=500$ control locations per used location for the parameter estimation.}
    \label{fig:ssf_ivs}
\end{figure}

To initialise the parameter estimation, we used the true parameter values and $50$ sets of randomly drawn parameter values as starting values. With the true values being the starting values, the estimated parameters did not diverge and the MS-iSSA performed well (Figure \ref{fig:ssf0}). This was, however, not the case when testing $50$ sets of random starting values (Figure \ref{fig:ssf_ivs}). While the estimates corresponding to state $1$ of the $2$-state MS-iSSA seem fine, the estimates of state $2$ completely diverged. This was especially the case for the shape and rate parameter of the gamma distribution.

\begin{figure}[t]
    \centering
    \includegraphics[width=0.5\textwidth]{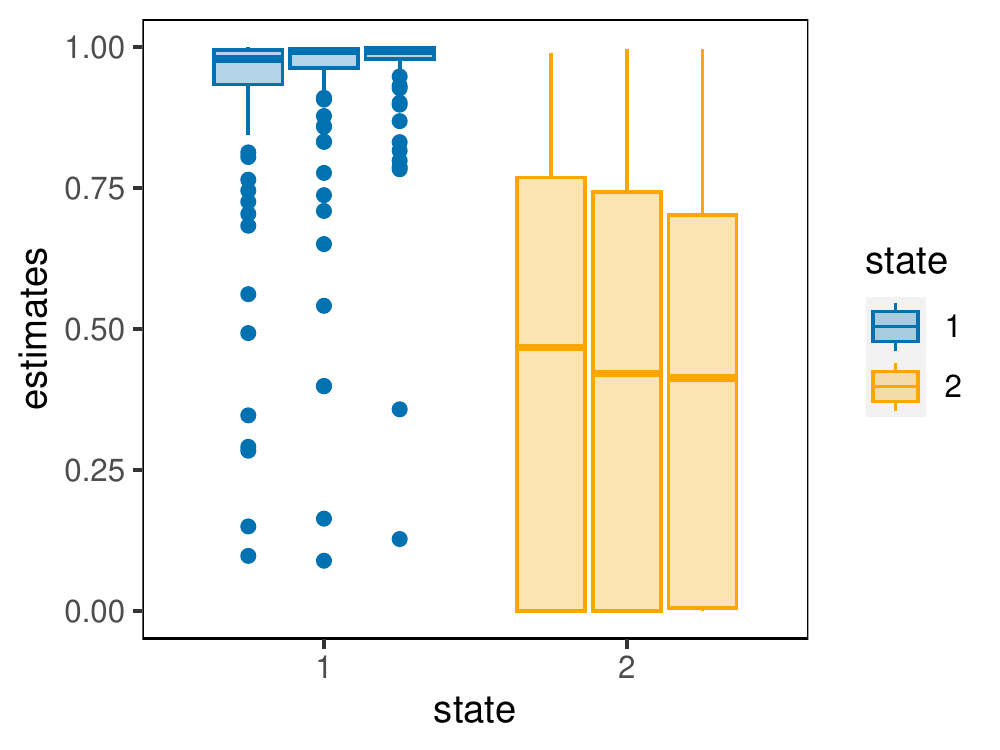}
    \caption{Boxplots of the estimated MS-iSSA transition probabilities $\gamma_{11}$ and $\gamma_{22}$, i.e.\ the probabilities to remain in the same state, across the $100$ simulation runs for the iSSA-scenario for each number of control locations $M$, respectively. Here, $50$ random starting values were used as starting values for the Markov-switching conditional logistic regression. The colours indicate the different states (state 1: blue, state 2: orange), the three adjacent boxplots refer the use of $M=20$, $M=100$ and $M=500$ control locations per used location for the parameter estimation.}
    \label{fig:ssf_gamma}
\end{figure}

This highlights the importance to carefully inspect the estimated MS-iSSA before interpreting the results. For this simulation scenario, there are several indications that the applied $2$-state MS-iSSAs are not appropriate for the simulated data sets, for example:
\begin{enumerate}
    \item Maximum likelihood values found across the $50$ sets of random starting values: 
    \item[] In some simulation runs only a single set of initial values led to the maximum likelihood value found, the remaining $49$ sets of initial values led to smaller log-likelihood values. This usually indicates numerical instability and problems in the estimated model.
    \item Estimated transition probabilities and Viterbi decoding:
    \item[] In some simulation runs the movement and selection coefficients of state $2$ differed between MS-iSSAs with similar log-likelihood values, i.e.\ different sets of starting values resulted in similar log-likelihood values, but different estimates for state $2$. This corresponds to problems in the estimated transition probability matrix, namely that the probability to stay in state $1$, $\gamma_{11}$, is (almost) equal to one, while the probability to stay in state $2$, $\gamma_{22}$, is low (Figure \ref{fig:ssf_gamma}). Consequently, it is possible that state $2$ is rarely or even never visited. This is also indicated by the Viterbi-decoded state sequences, which assigned less than $1\%$ of the data to state $2$ in $49\%$ ($M=20$), $46\%$ ($M=100$) and $58\%$ ($M=500$) of the simulation runs, respectively. Furthermore, in some simulation runs the probability to remain in the current state was low for both states, e.g.\ $\gamma_{11}<0.2$ for all states $i=1,\ldots,N$. This would correspond to a permanent switching between the states which is usually implausible in the context of animal movement data. 
    \item Estimated movement parameters:
    \item[] The shape, rate and concentration parameters must be greater than zero. In some simulation runs, however, they were estimated to be (almost) zero, which corresponds to their boundary values and indicates serious model problems.
    \item Variance of the state-dependent gamma distribution for step length:
    \item[] In some simulation runs, the variance of the state-dependent gamma distribution for state $2$, as derived from the estimated shape and rate parameters, was very low (below $0.1$). This implies that state $2$ only covers very specific step length values.
\end{enumerate}

Usually, several of such problems occur together. Furthermore, in this simulation scenario, the AIC performed worse in selecting the correct model (Table \ref{tab:ssf_ic}). Thus, this simulation scenario highlights the importance to test several sets of starting values to derive the maximum likelihood estimates and to closely inspect the model results. 

\begin{table}[ht]
\centering
\begin{tabular}[t]{lcccc}
\toprule
 & \multicolumn{2}{c}{AIC} & \multicolumn{2}{c}{BIC}\\\cmidrule(lr{.75em}){2-3}\cmidrule(lr{.75em}){4-5}
 no.\ cont.\ & iSSA & MS-iSSA & iSSA & MS-iSSA\\
\midrule
20 & \textbf{53} & 47 & \textbf{100} & 0\\
100 & \textbf{50} & 50 & \textbf{100} & 0\\
500 & \textbf{53} & 47 & \textbf{100} & 0\\
\bottomrule
\end{tabular}
\caption{Percentage of simulation runs in which either the iSSA or the MS-iSSA was selected by either AIC or BIC for each number of control steps used for model fitting, respectively. The cells belonging to the true underlying model are highlighted in bold face.}
\label{tab:ssf_ic}
\end{table}

\clearpage

\subsection{Habitat feature with less spatial variation}

\begin{figure}[h!t]
\includegraphics[width=\textwidth]{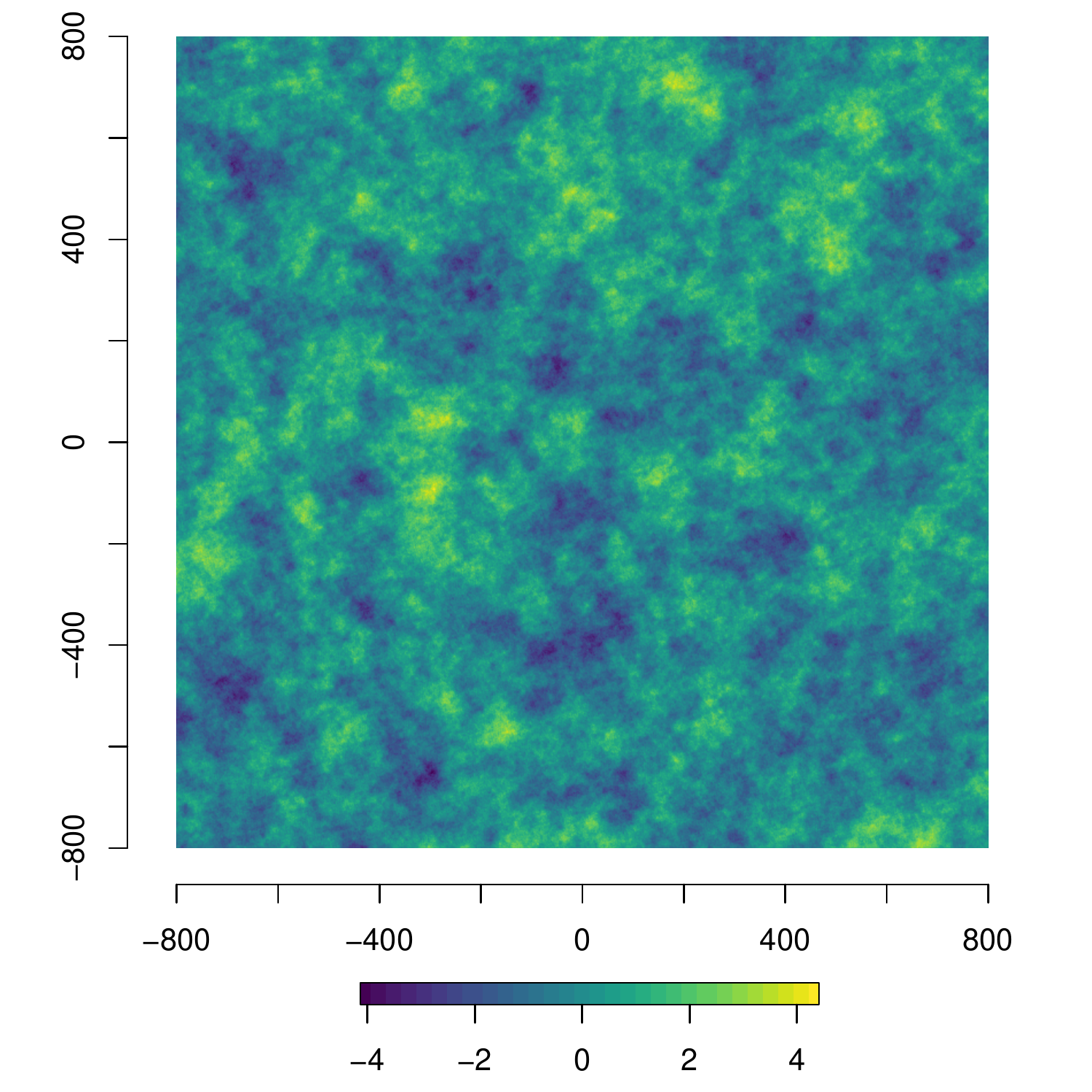}
\caption{Simulated landscape feature map used to derive the habitat covariates in the additional simulation study with less spatial variation in the habitat feature. The landscape feature map is a realisation of a Gaussian random field with covariance $1$ and range $50$.}
\label{fig:land50}
\end{figure}

\begin{figure}[ht]
\includegraphics[width=\textwidth]{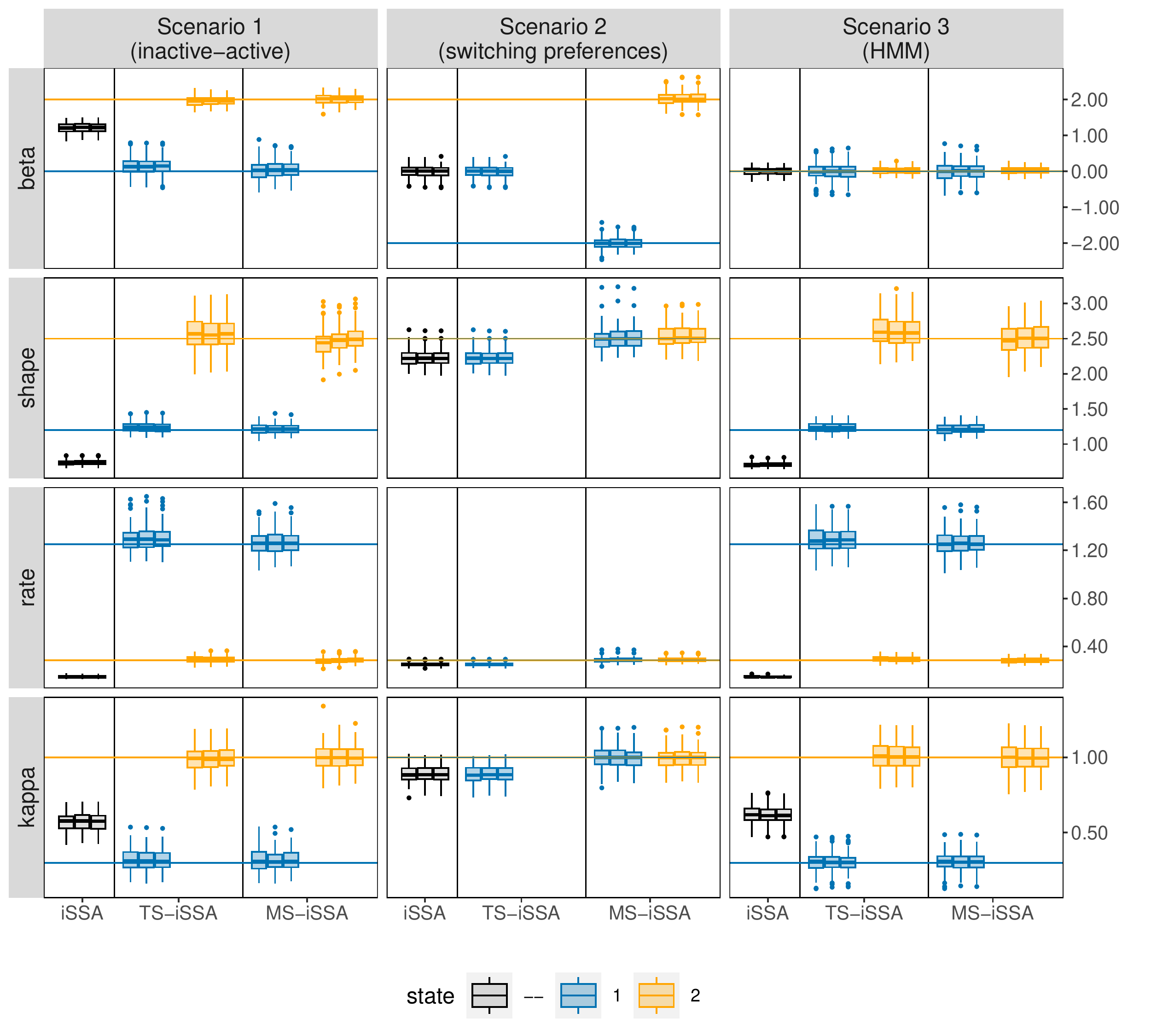}
\caption{Boxplots of the parameter estimates across the $100$ simulation runs in the simulation run with less spatial variation in the landscape feature. Results are shown for each applied method, simulation scenario and number of control locations $M$, respectively. The rows refer to the estimated selection coefficient (beta), the shape and rate of the gamma distribution for step length and the concentration parameter (kappa) of the von Mises distribution for turning angle, respectively. The columns refer to the three different simulation scenarios. For each method (iSSA, TS-iSSA and MS-iSSA) and state (state 1: blue, state 2: orange, no state differentiation: black), the three adjacent boxplots refer the use of $M=20$, $M=100$ and $M=500$ control locations per used location for the parameter estimation. Note that in Scenario 2, the TS-iSSA is naturally not capable to distinguish between two states as both share the same movement kernel. Thus, there are only results for a single state.}
\label{fig:Sim_res_l3}
\end{figure}

We re-ran the simulation study (Section 2.4, main manuscript) again using a landscape feature map with less spatial variation, i.e.\ a more homogeneous landscape (Figure \ref{fig:land50}). The results are similar to the results of our original simulation study, but the variance in the parameter estimates increased (Figure \ref{fig:Sim_res_l3}). Moreover, the classification performance of the HMM improves especially in Scenario 1 as with less spatial variation in the habitat features, the influence of the movement kernel on the movement track increases (Table \ref{tab_supp:Viterbi3}). This also leads to a decline in the classification performance of the MS-iSSA in Scenario 2.

\clearpage

\begin{table}[h!t]
\centering
\begin{tabular}[t]{lcccccccc}
\toprule
  & \multicolumn{2}{c}{MS-iSSA 20} & \multicolumn{2}{c}{MS-iSSA 100} & \multicolumn{2}{c}{MS-iSSA 500} & \multicolumn{2}{c}{HMM}\\
\midrule
scen. 1 & 2.83 & (0.68) & 2.73 & (0.68) & \textbf{2.70} & (0.64) & 2.98 & (0.65)\\
scen. 2 & 8.98 & (1.67) & 8.56 & (1.64) & \textbf{8.55} & (1.78) & 49.48 & (4.46)\\
scen. 3 & 2.48 & (0.49) & 2.42 & (0.51) & \textbf{2.37} & (0.53) & 2.39 & (0.53)\\
\bottomrule
\end{tabular}
\caption{Simulation runs with less spatial variation in the habitat feature: Mean missclassification rate with standard deviation in parentheses across the $100$ simulation runs for each scenario and fitted state-switching model, respectively. The missclassification rate is calculated as the percentage of states incorrectly classified using the Viterbi sequence. The lowest missclassification rate for each scenario is highlighted in bold face.}
\label{tab_supp:Viterbi3}
\end{table}

\clearpage

\section{Supplementary material for the bank vole case study}

\clearpage

\subsection{MS-iSSA estimates for interactions with same-sex conspecifics}
\begin{figure}[!h]
    \centering
    \includegraphics[width=0.99\textwidth]{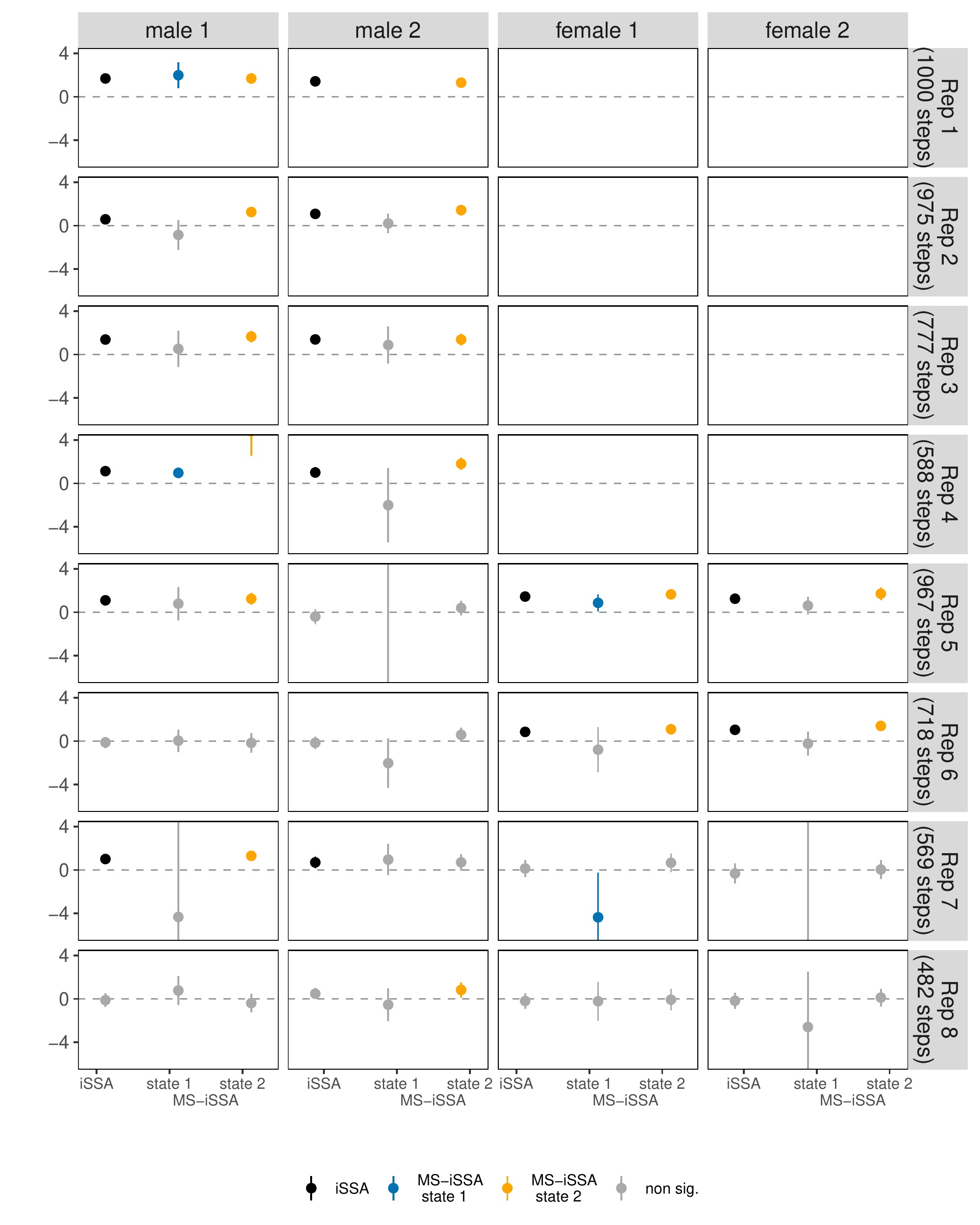}
    \caption{Estimated iSSA and MS-iSSA selection coefficients (solid points/triangles) of interaction behaviour between individuals of same sexes within the eight replicates (1–8), including $95\%$ confidence intervals (solid lines). Each replicate consisted of two males (male 1 and male 2) and one or two females (female 1 and female 2) such that each
    individual could respond to one same‐sex individual. Non‐significant coefficients (p-values below 0.05) are greyed out. The horizontal dashed line indicates zero (i.e. neutral behaviour); positive and negative coefficients indicate attraction and avoidance, respectively.}
    \label{fig:case_coeff_same}
\end{figure}

\clearpage

\subsection{TS-iSSA results}

\begin{figure}[ht]
    \centering
    \includegraphics[width=\textwidth]{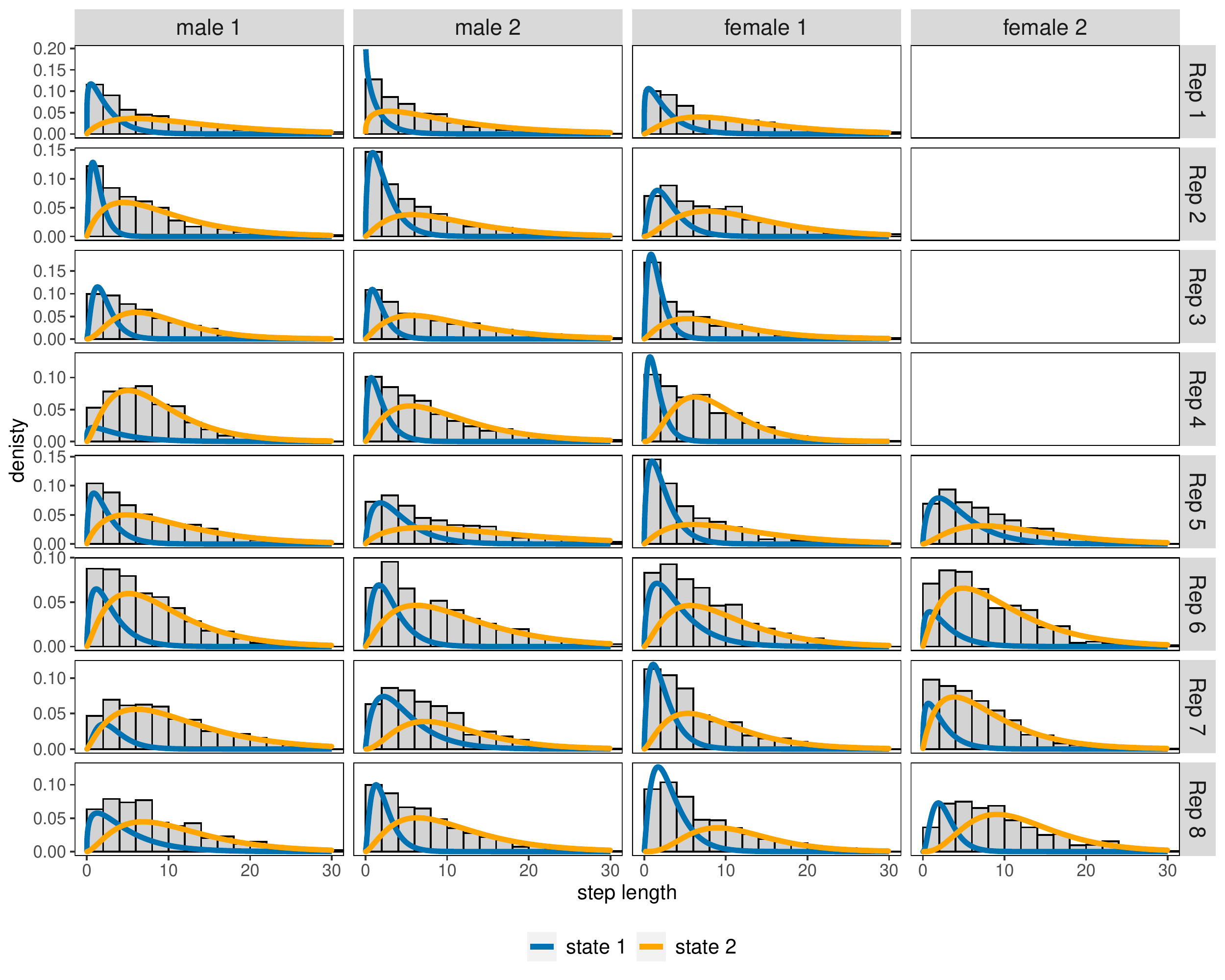}
    \caption{Estimated state-dependent gamma distributions for step length as implied by the fitted $2$-state TS-iSSAs for each individual in replicates $1-8$, respectively. The distributions are weighted by the relative state occupancy frequencies derived from the Viterbi sequence. The gray histograms in the background show the distribution of the observed step length.}
    \label{fig:gamma_ts}
\end{figure}

\clearpage

\begin{figure}[ht]
    \centering
    \includegraphics[width=\textwidth]{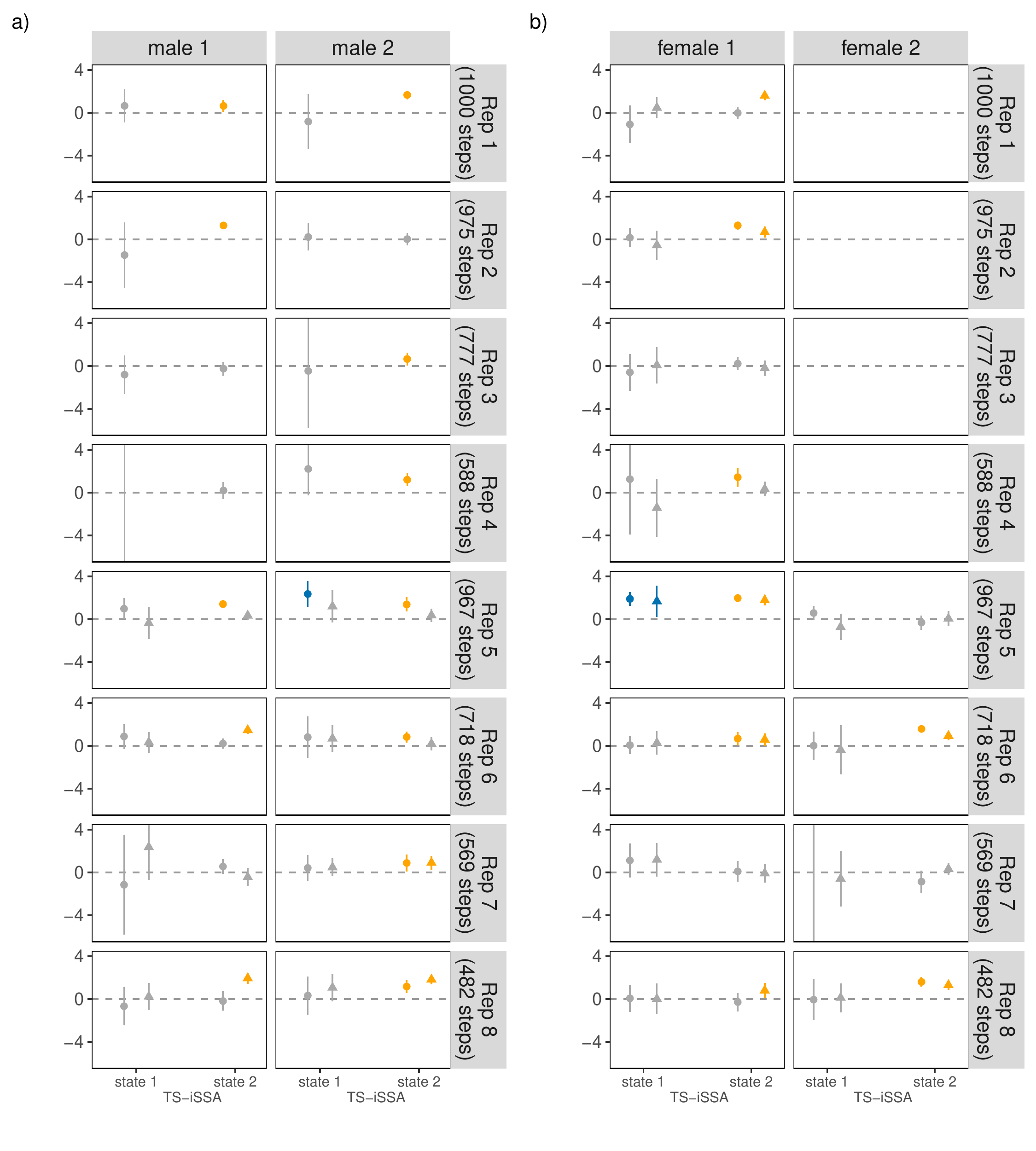}
    \caption{Estimated TS-iSSA selection coefficients (solid points/triangles) of interaction behaviour between individuals of opposing sexes within the eight replicates (1–8), including $95\%$ confidence intervals (solid lines). Each replicate consisted of two males (male 1 and male 2) and one or two females (female 1 and female 2) such that each
    individual could respond to up to two opposite‐sex individuals (dot: response to female/male 1, triangle: response to female/male 2 within a replicate). Non‐significant coefficients (p-values below 0.05) are greyed out. The horizontal dashed line indicates zero (i.e. neutral behaviour); positive coefficients indicate attraction, while negative
    coefficients would indicate avoidance}
    \label{fig:TS_coeff_same}
\end{figure}

\clearpage

\begin{figure}[ht]
    \centering
    \includegraphics[width=\textwidth]{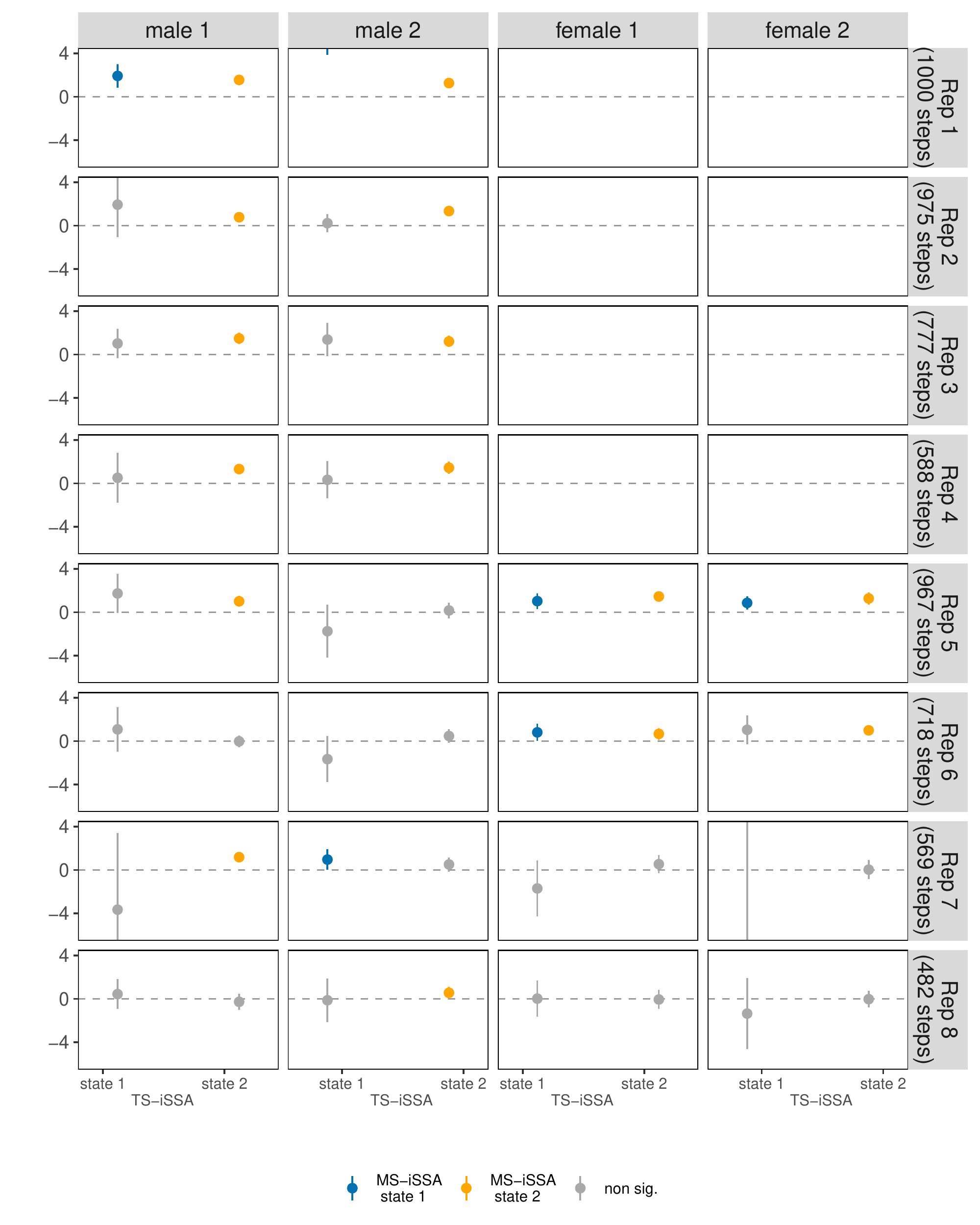}
    \caption{Estimated TS-iSSA selection coefficients (solid points/triangles) of interaction behaviour between individuals of same sexes within the eight replicates (1–8), including $95\%$ confidence intervals (solid lines). Each replicate consisted of two males (male 1 and male 2) and one or two females (female 1 and female 2) such that each
    individual could respond to one same‐sex individual. Non‐significant coefficients (p-values below 0.05) are greyed out. The horizontal dashed line indicates zero (i.e. neutral behaviour); positive coefficients indicate attraction, while negative
    coefficients would indicate avoidance.}
    \label{fig:case_coeff_mm}
\end{figure}

\clearpage

\subsection{Viterbi-decoding}

\begin{table}[ht]
\centering
\begin{tabular}[t]{clcccc}
\toprule
 & & \multicolumn{2}{c}{HMM} & \multicolumn{2}{c}{MS-iSSA} \\\cmidrule(lr{.75em}){3-4}\cmidrule(lr{.75em}){5-6}
Rep.\ & Ind.\ & State 1 & State 2 & State 1 & State 2\\\midrule
\multirow{3}{*}{1} & male 1 & 37.60 & 62.40 & 38.00 & 62.00\\
        & male 2 & 25.40 & 74.60 & 24.50 & 75.50\\
        & female 1 & 36.60 & 63.40 & 37.90 & 62.10\\\midrule
\multirow{3}{*}{2} & male 1 & 24.41 & 75.59 & 44.41 & 55.59\\
        & male 2 & 46.77 & 53.23 & 48.10 & 51.90\\
        & female 1 & 32.41 & 67.59 & 31.59 & 68.41\\\midrule
\multirow{3}{*}{3} & male 1 & 33.72 & 66.28 & 33.46 & 66.54\\
        & male 2 & 26.64 & 73.36 & 26.38 & 73.62\\
        & female 1 & 42.86 & 57.14 & 42.86 & 57.14\\\midrule
\multirow{3}{*}{4} & male 1 & 11.22 & 88.78 & 96.60 & 3.40\\
        & male 2 & 25.68 & 74.32 & 21.77 & 78.23\\
        & female 1 & 28.23 & 71.77 & 32.65 & 67.35\\\midrule
\multirow{3}{*}{5} & male 1 & 28.54 & 71.46 & 38.57 & 61.43\\
        & male 2 & 40.74 & 59.26 & 39.61 & 60.39\\
        & female 1 & 47.57 & 52.43 & 49.43 & 50.57\\
        & female 2 & 50.98 & 49.02 & 47.98 & 52.02\\\midrule
\multirow{3}{*}{6} & male 1 & 26.04 & 73.96 & 66.71 & 33.29\\
        & male 2 & 29.11 & 70.89 & 33.84 & 66.16\\
        & female 1 & 40.67 & 59.33 & 21.87 & 78.13\\
        & female 2 & 15.18 & 84.82 & 33.15 & 66.85\\\midrule
\multirow{3}{*}{7} & male 1 & 14.76 & 85.24 & 15.29 & 84.71\\
        & male 2 & 49.38 & 50.62 & 46.92 & 53.08\\
        & female 1 & 39.89 & 60.11 & 37.43 & 62.57\\
        & female 2 & 20.04 & 79.96 & 22.67 & 77.33\\\midrule
\multirow{3}{*}{8} & male 1 & 36.31 & 63.69 & 44.19 & 55.81\\        
        & male 2 & 30.08 & 69.92 & 45.85 & 54.15\\
        & female 1 & 54.98 & 45.02 & 56.64 & 43.36\\
        & female 2 & 24.27 & 75.73 & 24.48 & 75.52\\
\bottomrule
\end{tabular}
\caption{Percentage of observed steps decoded to belong to state 1 or state 2 according to the HMM (i.e.\ TS-iSSA) or MS-iSSA Viterbi sequence for each of the $28$ bank vole individuals.}
\end{table}

\clearpage

\subsection{Information criteria results}

\begin{table}[ht]
\centering
\footnotesize
\begin{tabular}[t]{cclcrcr}
\toprule
Rep.\ & No.\ steps & Ind.\ & AIC & $\Delta$AIC & BIC & $\Delta$BIC\\\midrule
\multirow{3}{*}{1} & \multirow{3}{*}{1000} & male 1 & MS-iSSA & 64.10 & MS-iSSA & 44.46\\
  & & male 2 & MS-iSSA & 120.51 & MS-iSSA & 100.88\\
  & & female 1 & MS-iSSA & 43.90 & MS-iSSA & 24.27\\\midrule
\multirow{3}{*}{2} & \multirow{3}{*}{975}  & male 1 & MS-iSSA & 79.87 & MS-iSSA & 60.34\\
  & & male 2 & MS-iSSA & 28.94 & MS-iSSA & 9.41\\
  & & female 1 & MS-iSSA & 49.66 & MS-iSSA & 30.13\\\midrule
\multirow{3}{*}{3} & \multirow{3}{*}{777}  & male 1 & MS-iSSA & 27.73 & MS-iSSA & 9.11\\
  & & male 2 & MS-iSSA & 23.31 & MS-iSSA & 4.69\\
  & & female 1 & HMM & 6.43 & HMM & 25.05\\\midrule
\multirow{3}{*}{4} & \multirow{3}{*}{588}  & male 1 & MS-iSSA & 23.62 & iSSA & 7.02\\
  & & male 2 & MS-iSSA & 47.20 & MS-iSSA & 29.70\\
  & & female 1 & MS-iSSA & 9.88 & HMM & 7.63\\\midrule
\multirow{4}{*}{5} & \multirow{3}{*}{967}  & male 1 & MS-iSSA & 83.41 & MS-iSSA & 54.17\\
  & & male 2 & MS-iSSA & 47.21 & MS-iSSA & 17.97\\
  & & female 1 & MS-iSSA & 239.92 & MS-iSSA & 210.68\\
  & & female 2 & MS-iSSA & 28.75 & HMM & 0.50\\\midrule
\multirow{4}{*}{6} & \multirow{3}{*}{718} & male 1 & MS-iSSA & 74.52 & MS-iSSA & 47.07\\
 & & male 2 & MS-iSSA & 13.93 & HMM & 13.53\\
 & & female 1 & MS-iSSA & 31.99 & MS-iSSA & 4.53\\
 & & female 2 & MS-iSSA & 72.20 & MS-iSSA & 35.59\\\midrule
\multirow{4}{*}{7} & \multirow{3}{*}{569} & male 1 & MS-iSSA & 27.22 & MS-iSSA & 1.16\\
  & & male 2 & MS-iSSA & 7.41 & HMM & 16.17\\
  & & female 1 & HMM & 0.54 & HMM & 26.61\\
  & & female 2 & MS-iSSA & 0.37 & HMM & 25.70\\\midrule
\multirow{4}{*}{8} & \multirow{3}{*}{482}  & male 1 & MS-iSSA & 46.04 & MS-iSSA & 20.97\\
  & & male 2 & MS-iSSA & 75.30 & MS-iSSA & 41.88\\
  & & female 1 & HMM & 7.41 & HMM & 32.48\\
  & & female 2 & MS-iSSA & 13.66 & iSSA & 19.76\\
\bottomrule
\end{tabular}
\caption{Models selected by either AIC or BIC for each of the $28$ bank vole individuals. The set of candidate models included the iSSA model, HMM and MS-iSSA model. $\Delta$ AIC and $\Delta$ BIC show the differences in AIC and BIC values between the best and second best model in the ranking.}
\label{tab:case_selection}
\end{table}

\clearpage

\bibliography{bibliography}\addcontentsline{toc}{section}{References}

\end{spacing}